\documentclass[english]{IEEEtran}
\usepackage{eqnarray,amsmath}
\usepackage[english]{babel}
\usepackage[utf8x]{inputenc}
\usepackage[T1]{fontenc}
\usepackage{float}
\usepackage{amssymb}
\usepackage{graphicx}
\usepackage{xcolor}
\usepackage{multirow}
\usepackage{adjustbox}
\usepackage{subfig}
\usepackage{babel}
\usepackage{array}
\usepackage{units}
\usepackage{mathdots}

\makeatletter

\providecommand{\tabularnewline}{\\}
\floatstyle{ruled}
\newfloat{algorithm}{tbp}{loa}
\providecommand{\algorithmname}{Algorithm}
\floatname{algorithm}{\protect\algorithmname}

\usepackage{stmaryrd}
\usepackage{blkarray}

\@ifundefined{showcaptionsetup}{}{%
 \PassOptionsToPackage{caption=false}{subfig}}
\usepackage{subfig}
\makeatother

\begin{document}
  \title{JPEG Steganography and Synchronization of DCT Coefficients\\ for a Given Development Pipeline}
  \author{Théo Taburet$^\times$, Patrick Bas$^\times$, Wadih Sawaya$^\sharp$,
  and Remi Cogranne$^+$\\
  $^\times$CNRS, Ecole Centrale de Lille, CRIStAL Lab, 59651 Villeneuve d\textquoteright Ascq Cedex, France,
  \{theo.taburet,patrick.bas\}@centralelille.fr\\
  $^\sharp$IMT Lille-Douai, Univ. Lille, CNRS, Centrale Lille, UMR 9189, France, wadih.sawaya@imt-lille-douai.fr\\
  $^+$ICD - M2S - ROSAS - FRE 2019 CNRS - Troyes University of Technology, Troyes, France, remi.cogranne@utt.fr}
  \maketitle

  \begin{abstract}
    This paper proposes to use the statistical analysis of the correlation between DCT coefficients to design a new synchronization strategy that can be used for cost-based steganographic schemes in the JPEG domain. First, an analysis is performed on the covariance matrix of DCT coefficients of neighboring blocks after a development similar to the one used to generate BossBase. This analysis exhibits groups of uncorrelated coefficients: 4 groups per block and 2 groups of uncorrelated diagonal neighbors together with groups of mutually correlated coefficients groups of 6 coefficients per blocs and 8 coefficients between 2 adjacent blocks. Using the uncorrelated groups, an embedding scheme can be designed using only 8 disjoint lattices. The cost map for each lattice is updated firstly by using an implicit underlying Gaussian distribution with a variance directly computed from the embedding costs, and secondly by deriving conditional distributions from multivariate distributions. The covariance matrix of these distributions takes into account both the correlations exhibited by the analysis of the covariance matrix and the variance derived from the cost. This synchronization scheme enables to obtain a gain of $P_E$ of at least $7\%$ at $QF95$ for an embedding rate close to 0.3 bnzac coefficient using DCTR feature sets for both UERD and JUniward.
\end{abstract}
%\NRC{Pas de parenthèse fermante + compliqué à comprendre dès l'abstract = simplifié en disant ``seulement'' qu'on peut clusteriser les coefficients de façon à mettre de côtés ceux qui sont correlées entre-eux mais non corrélées avec les autres}.

% \setcopyright{acmcopyright}
% \copyrightyear{2020}
% \acmYear{2020}
% \acmDOI{10.1145/1122445.1122456}

%% These commands are for a PROCEEDINGS abstract or paper.
% \acmConference[IH\&MMSEC '20]{IH\&MMSEC '20: ACM Information Hiding \& Multimedia Security}{June 22--24, 2020}{Denver, CO}
% \acmBooktitle{IH\&MMSEC '20: ACM Information Hiding \& Multimedia Security, June 22--24, 2020, Denver, CO}
% \acmPrice{15.00}
% \acmISBN{978-1-4503-XXXX-X/18/06}

% \begin{CCSXML}
% <ccs2012>
% <concept>
% <concept_id>10002978.10003022.10003028</concept_id>
% <concept_desc>Security and privacy~Domain-specific security and privacy architectures</concept_desc>
% <concept_significance>300</concept_significance>
% </concept>
% </ccs2012>
% \end{CCSXML}

%\ccsdesc[300]{Security and privacy~Domain-specific security and privacy architectures}
%\ccsdesc[500]{Security and privacy~Intrusion/anomaly detection and malware mitigation}
%\ccsdesc[500]{Security and privacy~Malware and its mitigation}

% \ccsdesc[300]{Security and privacy~Social network security and privacy}
% \ccsdesc[300]{Applied computing~Computer forensics}

%\keywords{Digital image steganography, JPEG domain, sensor noise, image processing pipeline, covariance}

%\renewcommand{\shortauthors}{The anonymous}

  %\maketitle

  \section{Introduction}

  % Need to take into account dependencies between samples
  % Cost based approaches 
  % CMD and Synch, lattice, gibbs sampling and more
  % Hybrid approaches Cost and embedding probabilities
  % Updist and Chinesse reference, in the DCT domain
  % Probabilistic approaches: Chain rule of embedding probabilities
  % Synchronization for Natural Steganography, pixel domain XX lattices, ICASSP and DCT domain 4*64 lattices, archiv
    \begin{figure*}[ht]
      \begin{centering}
      \subfloat[]{\begin{centering}
      \includegraphics[width=0.15\textwidth]{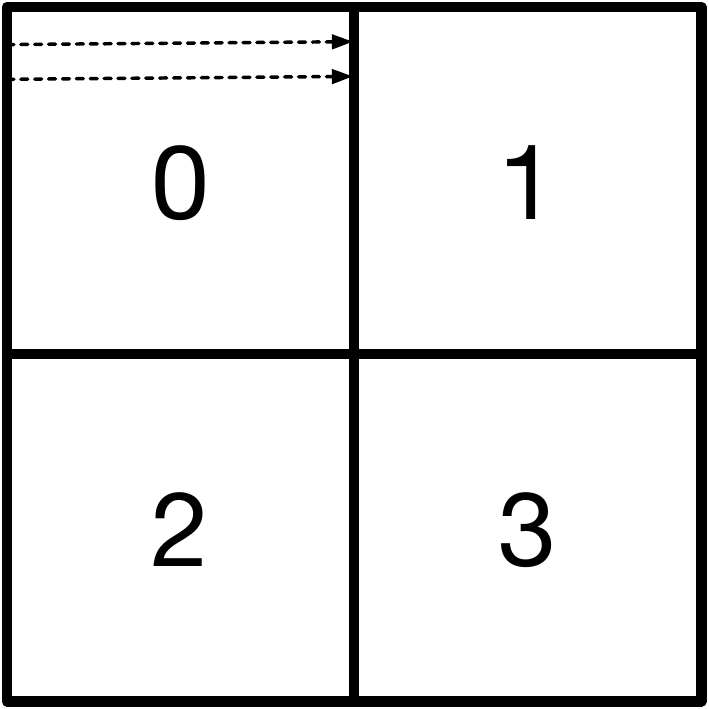}
      \par\end{centering}

      }\subfloat[]{\begin{centering}
      \includegraphics[width=0.6\textwidth]{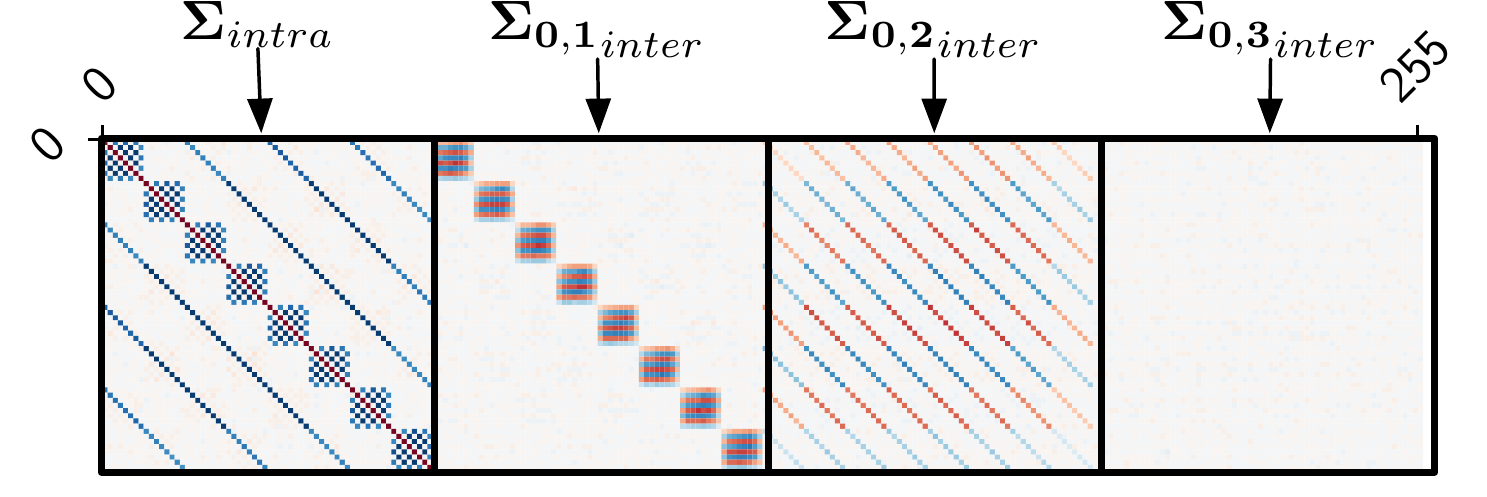}
      \par\end{centering}
      }

      \par\end{centering}
      \caption{(a) : scan order by block and coefficients, (b) Intra and inter correlations exhibited by the correlation matrix $\hat{\mathbf{R}}$. Blue colors denote negative correlation coefficients.\label{fig:scheme_corrs_2x2}}.
    \end{figure*}

  \subsection{Previous works}
    In order to increase the practical security of steganographic algorithms for digital images, one strategy is to synchronize embedding changes on samples that are correlated. The dependencies between image samples can come from correlations within the Cover contents, for example on homogeneous ares or textures, or correlations induced by the development pipeline (downscaling~\cite{bas:ICASSP-17}, demosaicking~\cite{tho2020natural}, DCT transforms~\cite{tho2020natural},...). 

    The synchronization process is, however, difficult to implement since this process is antagonist with the general principle of additive distortion commonly used in steganography~\cite{filler2011minimizing} which considers independent embedding changes and which is practically achieved using Syndrome Trellis Codes~\cite{filler2011minimizing}. 

    One common strategy to deal with this issue is to break the dependencies by decomposing the set of image coefficients (pixels or DCT coefficients) into sets of disjoint lattices. 

    Existing methods can be divided into two categories depending on whether synchronization if carried out on the cost-map or on the embedding probabilities.
    % It is possible to distinguish two classes of synchronization schemes, schemes synchronizing the cost map and scheme synchronizing embedding probabilities. \NRC{phrase avec beaucoup de répétition de ``synchro'' , je suggère : Existing methods can be divided into two categories depending on whether synchronization if carries out on  the cost-map or on the embedding probabilities.}

    {\bf Synchronization of the cost map}: The first scheme proposing to synchronize the cost map is based on Gibbs sampling, and it was proposed by Filler et al.~\cite{filler2010gibbs} and improved by Denemark et al.\cite{denemark2015improving} with the "synch" implementation. The proposed stego scheme works in the spatial domain and uses two lattices associated to a chessboard-like geometry: one the embedding is performed in the first lattice, the costs are adjusted in the second one so that consistent local modification changes are more likely performed. Independently, a very similar idea was proposed by Li et al~\cite{li2015strategy} using four lattices, but without performing multiple sweeps through the lattices (actually the analysis in~\cite{denemark2015improving} shows that only one sweep is necessary to maximize the performance, so the two strategies are very similar).

    {\bf Synchronization of embedding probabilities}: The other class of synchronization schemes proposes to modify the embedding probabilities directly, these schemes are dedicated to Natural Steganography proposed by Bas et al.~\cite{bas:WIFS-16}, where the stego signal tries to mimic the sensor photonic noise. 
    In order maximize the practical security after down-sampling~\cite{bas:ICASSP-17} in the spatial domain or demosaicking~\cite{tho2020natural} in the JPEG domain, the multivariate distribution of the stego signal is decomposed into conditional distributions overs disjoint lattices using the chain rule of conditional probabilities. 
    On a given lattice, the stego signal can be generated independently (conditionally to the embedding performed on the previous lattices) and a classical STC can be used. 
 
    Between these two classes there exist hybrid strategies proposed by Zhang et al.~\cite{zhang2016decomposing} and Li et al.~\cite{li2018defining} that define joint costs between samples and then derive a joint probability which is after decomposed into conditional probabilities and costs. 
 
    In the JPEG domain, the only schemes addressing this issues are proposed by Li et al.~\cite{li2018defining} and Taburet et al.~\cite{tho2020natural}. Even if these two schemes use completely different rationales and rely on completely different embedding schemes, they both try to preserve the continuities between adjacent JPEG blocks during embedding.
   
  \subsection{Main ideas}
    The present paper proposes a novel method that combines the advantages of both prior works~\cite{li2018defining,tho2020natural}. 
    On one hand, the method can be easily applied in practice in the sense that, as proposed in~\cite{li2018defining}, we use a classical JPEG embedding scheme cost map such as UERD~\cite{guo2015using} and J-UNIWARD~\cite{holub2014universal}. 
    On the other hand, the main contribution of the proposed method relies on its statistically-based foundation since, as in~\cite{tho2020natural}, it exploits the correlations induced by the development pipeline to synchronize the embedding changes. 
    However, contrary to~\cite{tho2020natural}, the proposed synchronization method can be applied with any cost based steganographic scheme.
    The main idea proposed in this paper is to leverage the natural correlations induced by the development pipeline to perform synchronization in the JPEG domain.
    % Similarly to~\cite{li2018defining}, this paper proposes to add a synchronization procedure to classical JPEG steganographic schemes based on cost maps such as UERD~\cite{guo2015using} and JUniward~\cite{holub2014universal}. Additionally, similarly to~\cite{tho2020natural}, the proposed scheme uses the knowledge of the correlations induced by the development pipeline to synchronize the embedding changes. However, contrary to~\cite{tho2020natural}, the proposed synchronization method can be applied on any cost based steganographic scheme.     

    % The main idea proposed in this paper is to leverage the natural correlations induced by the development pipeline to perform synchronization in the JPEG domain. 

    We first analyze the covariance matrix associated to a development in the DCT domain similar to the one performed to generate BossBase~\cite{bas:hal-00648057}, this is presented in section~\ref{sec:Analysis}. From this analysis, we are able to decompose the set of DCT coefficients into 8 disjoint lattices, in among lattice the different coefficients are mutually independent, see section~\ref{sec:synchro}.
    The embedding scheme is based on the conversion from the costs associated to each coefficients into an implicit zero-mean Gaussian distribution whose variance is directly computed from the cost. 
    This "Gaussian mapping", together with the Covariance matrix estimated in section~\ref{sec:Analysis} enables to compute a joint Gaussian distribution and to derive its associated conditional distributions which are used to performed synchronization on the 8 lattices. 
    The embedding scheme is presented in section~\ref{sec:embedding}. 
    Finally section~\ref{sec:results} presents the performance gains for different embedding strategies (UERD and JUniward) and different quality factors (QF 95), and analyze also the distribution of the payloads on the different lattices.

  % \subsection{Notations}

  %   Throughout this paper, we use capital letters for random variables and the corresponding lower-case symbols for their realizations. Matrices are typed in upper-case and vectors in lower-case boldface font. Matrix transpose will be denoted with a superscript "t" and we use following notations :
  %   \begin{itemize}
  %     \item $\hat{\mathbf{R}}$ the estimated correlation matrix, and $\mathbf{R}$ its thresholded version. $\mathbf{\Sigma}$ a covariance matrix.
  %     \item $C_{i}$ a continuous random variable producing $c_{i}$ the continous stego noise realisation and $\tilde{c_{i}}$ its discretized version.
  %     \item Hadamard product $\circ$ such that : $\mathbf{A}\circ\mathbf{B}=(a_{ij}\cdot b_{ij})\in \mathbb{R}^{n \times m}$ with $\left(\mathbf{A},\mathbf{B}\right)\in\mathbb{R}^{n\times m}\times\mathbb{R}^{n\times m}$
  %   \end{itemize}

\section{Correlations between DCT coefficients\label{sec:Analysis}}
  %Explain the development+[Display the 128x128] estimated covariance matrix after used for the development for BOSSBaseSameScale, show the threshold version, interpret intra [6 correlations per coefficient), interpret inter (8 correlations per coefficient). Decorrelated between diagonals. Compare the cov for other downscaling factors (to 256 instead of 512) and another non linear demosaicking algorithm.

  In this section we analyze the covariance matrix between DCT coefficients of neighboring $8\times 8$ DCT blocks after a development pipeline similar to the one used to generate BOSSBase (see Section~\ref{subsec:database} for more details on the development pipeline). Since the correlations related to the host content are difficult to model, we focused our analysis on the statistical model of the phonotic sensor noise. 

  We consequently computed the covariance matrix of $3 \times 3$  neighboring blocks of size $8 \times 8$ in the DCT domain (i.e. before quantization). The covariance matrix is estimated from 1000 constant-luminosity RAW images with photo-site values $\mu= 2^{12}$ coded on 14 bits and corrupted with an additive i.i.d. signal $S\sim\mathcal{N}(0,\,a\mu+b)$, demosaicked with the bi-linear algorithm, downsampled to a $512\times512$ images, and transformed into a 2D-DCT array. %The developed "image" in the DCT domain is then decomposed in $32\times32$ patches to derive the covariance matrix.%we estimate the correlation matrix $\hat{\mathbf{R}}$ between DCT coefficients of $3 \times 3$ neighboring of $8 \times 8$ blocks.

  In order to take into account symmetries of the whole covariance matrix (for example the fact that the covariance between two horizontal neighbors is identical, the analysis of only a portion of the covariance matrix can be conducted by considering only $2\times2$ adjacent blocks. It is illustrated Figure~\ref{fig:scheme_corrs_2x2}. The scan order for the four $8 \times 8$ DCT blocks consists of a scan by rows within each block and a block-wise scan across the four blocks as shown in Figure~\ref{fig:scheme_corrs_2x2}.

  By observing Figure~\ref{fig:scheme_corrs_2x2} together with the scan order and the decomposition of the matrix into different types, we can decompose the entire covariance matrix into four different types of $64 \times 64$ matrices : one intra-block covariance matrix and three inter-block covariance matrices:
  \begin{itemize}
    \item Intra-block $8 \times 8$ covariance matrices $\mathbf{R}_{i, i}$ capture the correlations between DCT coefficients of the same block. %They are located on the diagonal of the correlation matrix $\mathbf{R}$.
    \item Horizontal and vertical capture correlations between horizontal blocks and vertical blocks respectively. 
    \item Diagonal inter-block capture correlation between diagonal blocks.
  \end{itemize}

    % \begin{figure*}[h]
    %   \begin{centering}
    %   \subfloat[]{\begin{centering}
    %   \includegraphics[width=0.75\columnwidth]{Figures/cov_INTRAS_LIN_512x512_2x2}
    %   \par\end{centering}
    %   }

    %   \par

    %   \subfloat[]{\begin{centering}
    %   \includegraphics[width=0.25\columnwidth]{Figures/scheme_cov_2x2/scan_corr_2x2}
    %   \par\end{centering}

    %   }\subfloat[]{\begin{centering}
    %   \includegraphics[width=0.25\columnwidth]{Figures/scheme_cov_2x2/scheme_corr_2x2.pdf}
    %   \par\end{centering}
    %   }

    %   \par\end{centering}
    %   \caption{(a) Estimated correlation matrix $\hat{\mathbf{R}}$, (b) : scan order by block and coefficients, (c) : types of sub-matrices representing exhibited in the inter-block correlation matrix in (a).}
    % \end{figure*}

    Important remarks can be highlighted from the analysis of this convariance matrix:
    \begin{itemize}
      \item it is sparse, i.e. lot of DCT coefficients are uncorrelated, in one block one coefficient is correlated with 6  other ones, and on vertically or horizontally adjacent blocks one coefficient is correlated with 8 other ones in each block.
      \item two diagonal blocks are close to uncorrelated (the correlation values are very low).
      \item the patterns of the covariance matrix are immune to the type of demosaicking or down-sampling kernel. We tested the different demosaicking algorithm offered by the "rawpy" library and different down-sampling kernels and the patterns (but not the correlation values) are similar.
    \end{itemize}

In order to be invariance to sensor noise power (which is both dependent of the ISO setting of the camera and the sensor model), we convert the covariance matrix into a correlation matrix, were each diagonal terms equals 1, and each of diagonal term is divided by $\sigma_i \sigma_j$. Each term of the empirical covariance matrix is defined as: 

\begin{equation}
  \Sigma_{i,j}=\frac{1}{N}\sum_{k=1}^{N} (C_i(k)-\bar{C}_i])(C_j(k)-\bar{C}_j]),
\end{equation}
where $\bar{C}_i$ is the empirical mean of coefficient $C_i$, and each term of the matrix of correlation is consequently defined as: 

\begin{equation}
  \rho_{i,j}=\frac{\Sigma_{i,j}}{\sqrt{\Sigma_{i,i}\Sigma_{j,j}}}.
\end{equation}

The obtained matrix of correlation coefficients is presented in Figure~\ref{fig:scheme_corrs_2x2} and it appears that 8 modes are enough in order to describe the intra-block correlations while 6 modes by blocks are needed for the inter-block correlations. The others values being very low, we have thus decided to neglect them and to set them to zero.

\section{Lattice decomposition\label{sec:synchro}}
From these observations we can now decompose the set of DCT coefficients into lattices. Each lattice must be composed of uncorrelated coefficient. We end up with 8 lattices, because of the following observations:

- To deal  with intra-block correlations, we notice that we can find 4 groups of coefficients uncorrelated to one another. The 4 subsets (lattices) $\mathbf{\Lambda_i} \in \mathbb{N}^{16}$ with $i\in\left\{ 0,\ldots,3\right\}  $ of these mutually decorated modes indexes are arranged thanks to  a permutation matrix $\mathbf{P}$ such that :

  $$\mathbf{R}_{intra}=\mathbf{P}
  \underbrace{
  \left[
  \begin{array}{cccc}
  \mathbf{I_{16}} & \mathbf{\Sigma_{\Lambda_{0},\Lambda_{1}}} & \cdots & \mathbf{\Sigma_{\Lambda_{0},\Lambda_{3}}}\\
  \mathbf{\Sigma_{\Lambda_{1},\Lambda_{0}}} & \mathbf{I_{16}} & \ddots & \vdots\\
  \vdots & \ddots & \mathbf{I_{16}} & \Sigma_{\Lambda_{2},\Lambda_{3}}\\
  \mathbf{\Sigma_{\Lambda_{3},\Lambda_{0}}} & \cdots & \mathbf{\Sigma_{\Lambda_{3},\Lambda_{2}}} & \mathbf{I_{16}}
  \end{array}
  \right]
  }_{\mathbf{R_{p}}^{intra}}\mathbf{P}^{-1}$$

\begin{figure}[ht]
      \begin{centering}
      \includegraphics[width=0.7\columnwidth]{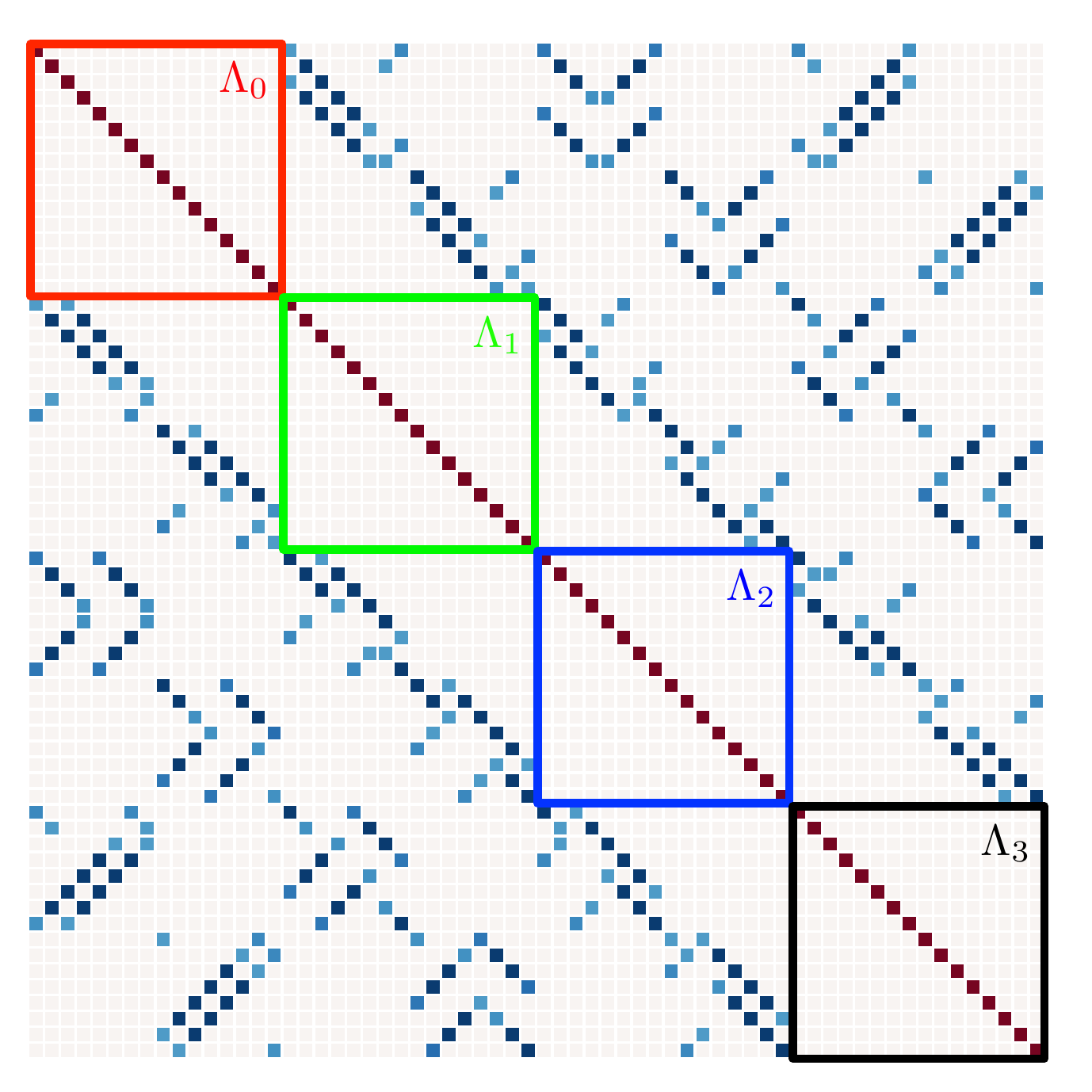}
      \par\end{centering}
      \caption{Intra-correlations matrix after permutation $\mathbf{R_{p}}^{intra}$ and associated lattices. \label{fig:scheme_lattices_intras_and_permutations}}
  \end{figure}

  The displayed correlation matrix ~\ref{fig:scheme_lattices_intras_and_permutations} after permutation of the indexes highlights the fact that a coefficient belonging to $\mathbf{\Lambda_0}$ will not depend on any previous realizations. However, we also notice that a coefficient belonging to $\mathbf{\Lambda_i}$ with $0<i<4$, depends on two coefficients of each of the lattices that precedes it.

  % For a given mode $m$ its matching column index $i_m$ in $\mathbf{R_{p}}^{intra}$ is number such that $\mathbf{P}_{m,\,i_{m}}^{t}=1$. Thus the set of mode indexes correlated with $i_{m}$ in $\mathbf{R_{p}}^{intra}$ can be obtained by :

  % $\mathrm{idx}^{intra}_{p}\left(i_{m}\right)=\left\{ i_{u}\mid\left(\mathbf{R_{p}}^{intra}\right)_{i_{u},\,j}=1,\:0\leq j<i_{m}\right\}$

  % Finally, the set of modes correlated with $m$ is given by :

  % $$\mathrm{idx}^{intra}\left(m\right)=\left\{ i_{v}\mid\mathbf{P}_{i_{v},\,i_{u}}=1,\:i_{u}\in\mathrm{idx}^{intra}_{p}\left(i_{m}\right),i_{m}\mid\mathbf{P}_{m,\,i_{m}}^{t}=1\right\}$$

  % For instance the mode $m=11 \in \mathbf{\Lambda_1}$ which row index is $17$ in $\mathbf{R_{p}}^{intra}$ is correlated with the modes $\left\{ 8,\,24\right\}$, $m=4 \in \mathbf{\Lambda_2}$ is correlated with $\left\{ 0,\,32,\,55,\,2\right\}$ and so on.

  - To deal with inter-block correlations, we proceed in the same way. This time, we can see from the analysis of the covariance matrix that each mode will then depends of 8 modes for each connected block (see. Figure ~\ref{fig:scheme_corrs_2x2}). We also notice that since two diagonally-connected block are uncorrelated, we can build 2 sub-lattices of $8 \times 8$ blocks to deal inter-block correlations. 

  % We can define $\mathrm{idx}^{inter}\left(m\right)$ the set of modes index correlated with $m$ within its blocks and its 1st order blocks neighborhood as :
  % $$\mathrm{idx}^{inter}\left(m\right)=\left\{ i_{v}\mid\left(\mathbf{R}^{inter}\right)_{i_{v},\,m+256},\:0\leq i_{v}<256\right\}\cup\mathrm{idx}^{intra}\left(m\right)$$

  % The DCT mode $m =2$ is correlated with the modes (as depicted on Figure ~\ref{fig:scheme_1st_order_corrs}):
  % \begin{itemize}
  %   \item $\left\{ 2,10,18,26,34,42,50,58\right\}$ from the upper and lower block,
  %   \item $\left\{ 0,1,2,3,4,5,6,7\right\}$ from the left and right block.
  %   \item $\left\{ 0,16\right\}$ within its block.
  % \end{itemize}

  \begin{figure}[ht]
      \begin{centering}
      \subfloat[]{\begin{centering}
      \includegraphics[width=0.5\columnwidth]{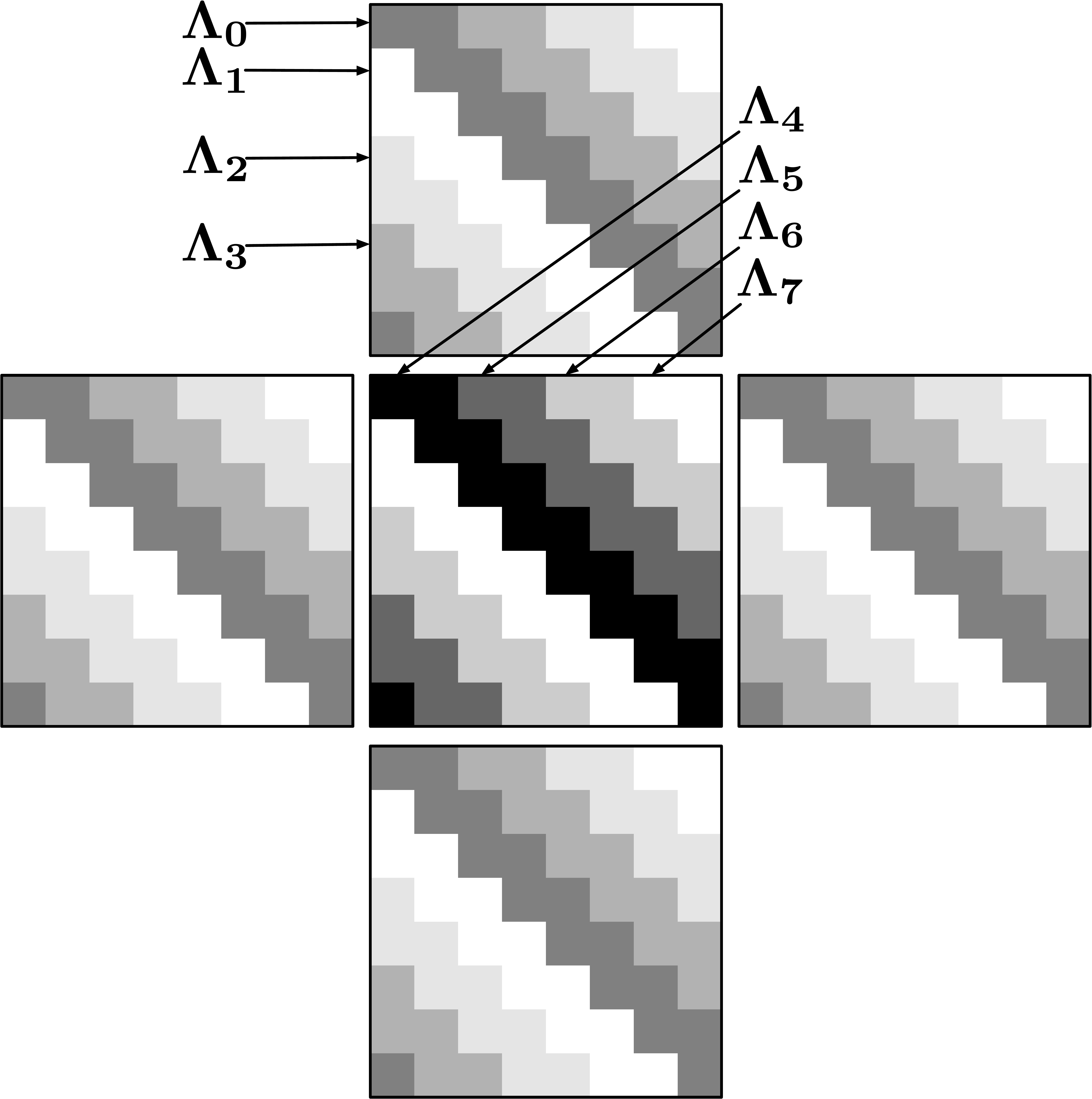}
      \par\end{centering}

      }\subfloat[]{\begin{centering}
      \includegraphics[width=0.5\columnwidth]{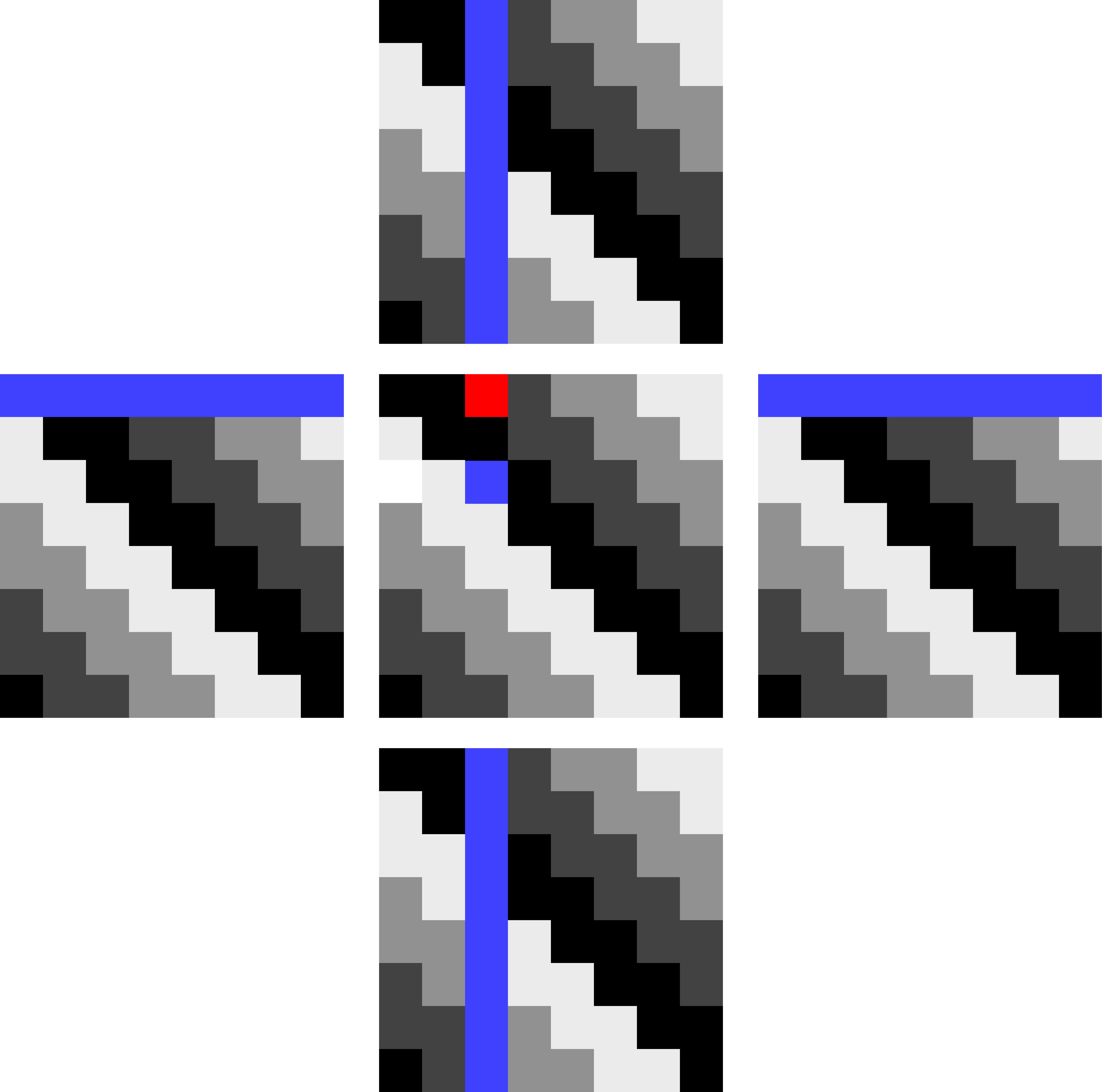}
      \par\end{centering}

      }
      \par\end{centering}
      \caption{(a) Decomposition of the DCT modes into 8 lattices, (b) Modes used to compute the conditional probability (blue) of mode $(0,2)\in \Lambda_5$ (red).\label{fig:scheme_1st_order_corrs}}
    \end{figure}
    \begin{table}[ht]
      \begin{centering}
      \begin{tabular}{|c|c|c|c|c|c|c|c|}
      \hline
      $\mathbf{\Lambda}_{0}$ & $\mathbf{\Lambda}_{1}$ & $\mathbf{\Lambda}_{2}$ & $\mathbf{\Lambda}_{3}$ & $\mathbf{\Lambda}_{4}$ & $\mathbf{\Lambda}_{5}$ & $\mathbf{\Lambda}_{6}$ & $\mathbf{\Lambda}_{7}$\\
      \hline 
      0 & 2 & 4 & 6 & 32 & 34 & 36 & 38\\
      \hline 
      \end{tabular}
      \par\end{centering}
      \caption{Number of correlated coefficients for each lattice.\label{tab:tab_correlated_coeffs}}
    \end{table}

Based on the above considerations, each image can be split into 8 disjoint lattices in order to sample a stego signal in the DCT domain preserving both intra-block and inter-block correlations. 

Figure~\ref{fig:scheme_1st_order_corrs} (a) shows the locations of the uncorrelated coefficients for the different lattices, and Figure~\ref{fig:scheme_1st_order_corrs} (b) highlights the locations of correlated coefficients belonging to previous lattices for one given mode. 

\begin{figure}[h]
      \begin{centering}
      \includegraphics[width=0.5\columnwidth]{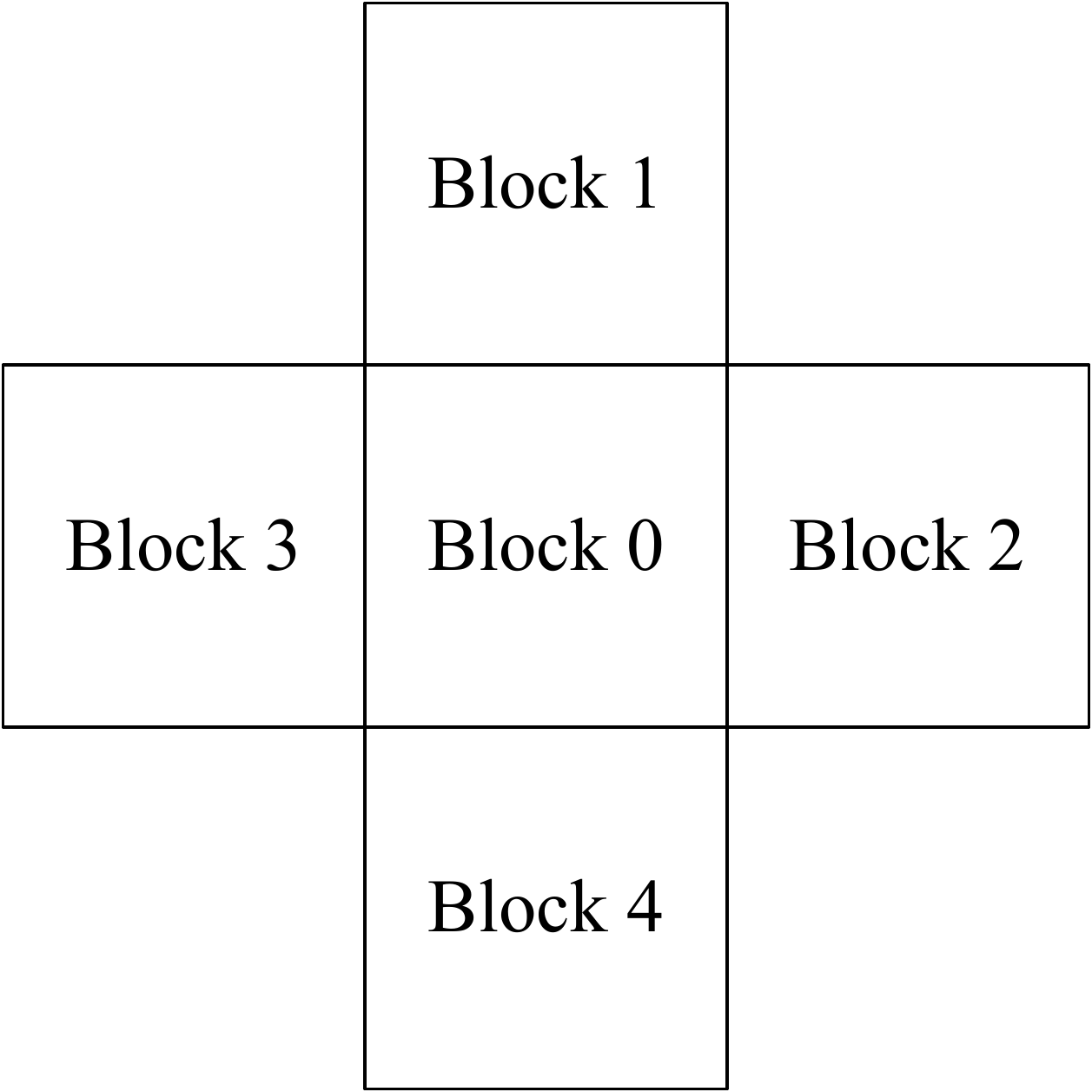}
      \par
      \end{centering}
      \caption{Block naming convention.\label{fig:blockOrder}}
    \end{figure}

Table \ref{tab:tab_correlated_coeffs} indicates for lattice $\mathbf{\Lambda_i}$ the number of correlated coefficients lattices $\{\mathbf{\Lambda_{i-1}},\dots, \mathbf{\Lambda_{0}}\}$ and Tables~\ref{tab_L1}, ~\ref{tab_L2}, ~\ref{tab_L3}, ~\ref{tab_L4}, ~\ref{tab_L5}, ~\ref{tab_L6} and ~\ref{tab_L7} (see the Appendices) exhibit for each mode of each lattices the different correlated modes belonging to previous lattices for the same block or adjacent ones as depicted on Figure~\ref{fig:blockOrder}.

\section{Embedding scheme\label{sec:embedding}}
  \begin{figure*}[t]
    \begin{centering}
    \includegraphics[width=1.9\columnwidth]{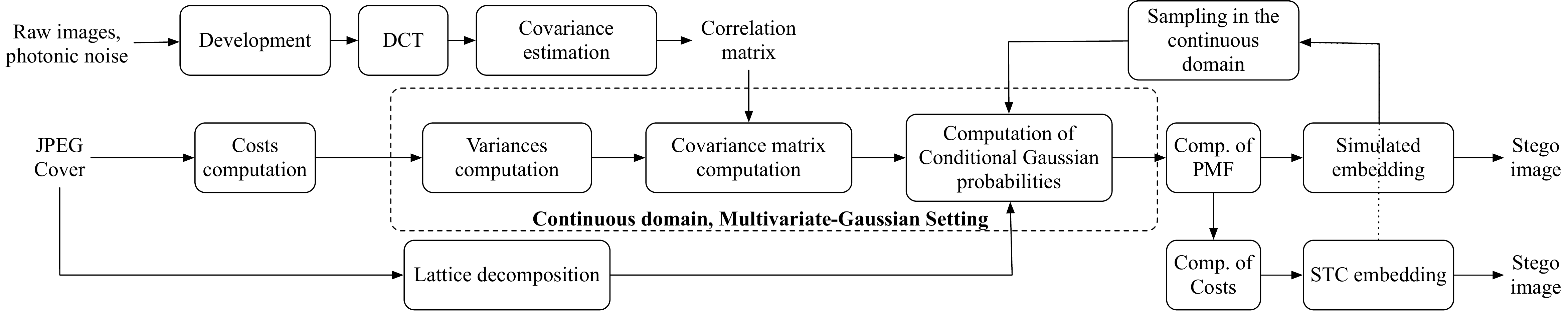}
    \par
    \end{centering}
    \caption{Overview of the embedding scheme.\label{fig:overview}}
  \end{figure*}

  We detail now how we can leverage the both the covariance matrix presented in section~\ref{sec:Analysis} and the lattice decomposition presented in section~\ref{sec:synchro} to enable to synchronization of embedding changes for cost-based embedding schemes.

  Figure~\ref{fig:overview} summarizes the different steps necessary to perform embedding. 

  The key idea of the proposed scheme is the computation of continuous Gaussian densities from the costs derived from the additive steganographic scheme. This setting can be justified by the fact that in order to leverage the covariance matrix of the sensor noise, we need to model the stego signal by Gaussian distribution since it is the only distribution that can be defined only by its expectation and its covariance. The derivation of variances from costs is detailed in section~\ref{subsec-cost-var}. 

  The next step is the construction of a covariance matrix reflecting the correlations coming from the development pipeline, it is detailled in section~\ref{subsec:emb-cov}.

  This covariance matrix, together with the history of the embedding changes performed on the previous lattices, are used to derive conditional Gaussian densities, which are in turn converted into a pmf (probability mass function) to simulate embedding, or cost to use STC embedding. This is detailled in section~\ref{subsec:emb-prob}.

  % \begin{algorithm}[h]
  %   - Divide the image into 8 macro-lattices $\Lambda_{i},\:i \in \left\{ 0,\ldots,7\right\} $;

  %   \begin{itemize}
  %     \item \textbf{For} each macro-lattice $\Lambda_{i}$ \textbf{do}:
  %     \begin{itemize}
  %       \item \textbf{For} each DCT block $\mathbf{x}$ of $\Lambda_{i}$ \textbf{do}:
  %       \begin{itemize}
  %         \item Compute the covariance matrix for each DCT block~Eq. (\ref{eq:Vect_DCT_2D_from_RAW});
  %         \item Compute the conditional mean vector (Eq.~(\ref{eq:conditional_mean}))
  %         and covariance matrix (Eq.~(\ref{eq:conditional_cov})) w.r.t. the
  %         embeddings done on the previous lattices;
  %       \begin{itemize}
  %         \item \textbf{For} each DCT coefficient \textbf{do}:
  %       \begin{itemize}
  %         \item Compute the conditional distribution Eq.~(\ref{eq:Cond_distrib});
  %         \item Compute the PMF $\pi_{i}(k)$, Eq. ~(\ref{eq:ERF});
  %         \item Perform the modification;
  %         \item Sample the continuous variable related to the modification, Eq~(\ref{eq:RejecSampling});
  %       \end{itemize}
  %       \end{itemize}
  %       \end{itemize}
  %     \end{itemize}
  %   \end{itemize}
  %   \caption{J-Cov-NS embedding scheme.\label{alg:Embedding-scheme.}}
  % \end{algorithm}

  \subsection{From costs to Gaussian distributions\label{subsec-cost-var}}
    Without loss of generality, we assume also ternary embedding. For a coefficient of coordinates $\left(i,j\right)$ into a $8\times8$ DCT block, we assume that the underlying unquantized stego signal is associated with an Normal distribution with zero mean and a variance $\sigma_{i,j}^{2}$. The variance is determined w.r.t the costs computed by an heuristic algorithm (UERD or J-UNIWARD here) for a given message size $\mathbf{m}$. 

    $$S_{i,j}\sim\mathcal{N}\left(0,\sigma_{i,j}^{2}\right).$$

    %No modifications of the coefficient ($\tilde{s_{i,j}}=0$) of coordinates $\left(i,j\right)$ stand for $s_{i,j}\in\left[-0.5,0.5\right]$, while $\tilde{s_{i,j}}=-1\Longleftrightarrow s_{i,j}\in\left[-\infty,-0.5\right]$ and $\tilde{s_{i,j}}=+1\Longleftrightarrow s_{i,j}\in\left[0.5,+\infty\right]$.

    For each coefficient $\left(i,j\right)$ we can compute the triplet of costs $( c_{i,j}^{-1}, c_{i,j}^{0}, c_{i,j}^{+1})$ respectively associated to the embedding changes ${-1, 0 ,+1}$. Since we use non side-informed scheme, we also assume that $ c_{i,j}^{-1}= c_{i,j}^{+1}$. 

    We can convert the costs into embedding probabilities using Lagrangian optimization~\cite{filler2011minimizing} by using the formula:
    \begin{equation}
P_{i,j}(k)=\frac{\exp\left(-\lambda c_{i,j}^{k}\right)}{\exp\left(-\lambda c_{i,j}^{0}\right)+\exp\left(-\lambda c_{i,j}^{+1}\right)+\exp\left(-\lambda c_{i,j}^{-1}\right)},
    \label{cost-probability}
    \end{equation}
    with $k\in{-1,0,+1}$, and $\lambda$ following the payload constraint.

    Denoting $q_{i,j}$ the JPEG quantization step associated to coefficient $(i,j)$,  we now assume that the embedding probabilities correspond to the probabilities of a quantized Gaussian distribution using three quantization bins, respectively $]-\infty,-q_{i,j}/2]$, $]-q_{i,j}/2,q_{i,j}/2]$, $]q_{i,j}/2,+\infty]$ for $-1, 0, +1$. The relation between $\sigma^2$ and the embedding probabilities is then given by:

%     We set the variance of the Gaussian distribution $\sigma_{i,j}^{2}$ to verify:

% \begin{equation}
%       \left(
%       \begin{array}{c}
%         \mathrm{P}_{i,j}\left(-1\right)\\
%         \mathrm{P}_{i,j}\left(0\right)\\
%         \mathrm{P}_{i,j}\left(+1\right)
%       \end{array}
%       \right)=\left(
%       \begin{array}{c}
%         \mathrm{P}\left(-\infty<S_{i,j}\leq-q_{i,j}/2\right)\\
%         \mathrm{P}\left(-q_{i,j}/2<S_{i,j}<q_{i,j}/2\right)\\
%         \mathrm{P}\left(q_{i,j}/2\leq S_{i,j}<\infty\right)
%       \end{array}
%       \right).
%       \label{eq:P_from_rho}
% \end{equation}

    % This estimation can be performed in two steps:
    % \begin{enumerate}
    %   \item \textbf{Derive embedding probabilities} $\left(P_{i,j}\left(-1\right),P_{i,j}\left(0\right),P_{i,j}\left(+1\right)\right)$ from $\left( c_{i,j}^{-1}, c_{i,j}^{0}, c_{i,j}^{+1}\right)$.

    %   \item \textbf{Find the corresponding Normal PDF} from : \\$\left(P_{i,j}\left(-1\right),P_{i,j}\left(0\right),P_{i,j}\left(+1\right)\right)$ with $s_{i,j}\sim\mathcal{N}\left(0,\sigma_{i,j}^{2}\right)$ and :

      \begin{equation}
      \sigma^2_{i,j}=\frac{q^2_{i,j}}{8\left(\mathrm{erf^{-1}}\left(-P_{i,j}\left(0\right)\right)\right)^2}.
      \label{eq:variance}
      \end{equation}

      % Such that :

      % $$
      % \begin{array}{c}
      %   P_{i,j}\left(-1\right)=P_{i,j}\left(+1\right)=\frac{1}{2}\left[1+\mathrm{erf}\left(\frac{-0.5}{\sigma\sqrt{2}}\right)\right]\\
      %   P_{i,j}\left(0\right)=1-P_{i,j}\left(-1\right)-P_{i,j}\left(+1\right)=-\mathrm{erf}\left(\frac{-0.5}{\sigma\sqrt{2}}\right)
      % \end{array}
      % $$
    %\end{enumerate}

    %It is thus possible to create $\mathbf{S_{\sigma}}$ the array whose each coefficient has the corresponding $\sigma^{2}$ value associated to its corresponding probabilities of modification.

    % 1. Derive embedding probabilities \left[P_{i,j}\left(-1\right),P_{i,j}\left(0\right),P_{i,j}\left(+1\right)\right] from costs \left[C_{i,j}\left(-1\right),C_{i,j}\left(0\right),C_{i,j}\left(+1\right)\right].

    % 2. Find the corresponding Normal PDF from \left[P_{i,j}\left(-1\right),P_{i,j}\left(0\right),P_{i,j}\left(+1\right)\right] such that c_{i,j}\sim\mathcal{N}\left(0,\sigma_{i,j}^{2}\right).

    % P_{i,j}\left(-1\right)=P_{i,j}\left(+1\right)=\frac{1}{2}\left[1+\mathrm{erf}\left(\frac{-0.5}{\sigma\sqrt{2}}\right)\right]

    % P_{i,j}\left(0\right)=1-P_{i,j}\left(-1\right)-P_{i,j}\left(+1\right)=-\mathrm{erf}\left(\frac{-0.5}{\sigma\sqrt{2}}\right)So from P_{i,j}\left(0\right) it is possible to derive an expression of \sigma_{i,j} :

    % \sigma_{i,j}=\frac{1}{\sqrt{2}\cdot\mathrm{erf^{-1}}\left(P_{i,j}\left(0\right)\right)}.
  \subsection{Construction of the covariance matrix\label{subsec:emb-cov}}
  %-------------------------------------------------------------------------

  The covariance matrix is built in  order to take into account the embedding changes that have already been made during the embedding. Its diagonal terms are given by (\ref{eq:variance}) and its off diagonal terms take into account the correlation coefficients $\rho_{k,l}$ estimated in section \ref{sec:Analysis}.

    For a given mode, the covariance matrix is built using the variances of the $(m-1)$ correlated coefficients and weighting the inter-correlations $\sigma_i\sigma_j$ such that their correlation coefficient equal the one estimated in section \ref{sec:Analysis}. Note that uncorrelated coefficients are not taken into account since they can be modified independently during the embedding. The resulting covariance matrix $\mathbf{\Sigma}_m$ is given by:

    \begin{equation}
    \mathbf{\Sigma}_m=\left[\begin{array}{ccccc}
      \sigma^2_1 & \rho_{1,2}\sigma_1 \sigma_2 & \cdots &  & \rho_{1,m}\sigma_1 \sigma_m\\
      \rho_{1,2}\sigma_1 \sigma_2 & \sigma^2_2\\
      \vdots &  & \ddots &  & \vdots\\
       &  &  & \sigma^2_{m-1} & \\
      \rho_{1,m} \sigma_1 \sigma_m &  & \cdots &  & \sigma^2_{m}
      \end{array}\right].
    \end{equation}

\subsection{Computation of embedding probabilities and costs\label{subsec:emb-prob}}
Once the covariance matrix is computed, we can derive the conditional pdf $P(c_m|c_{m-1},\dots,c_{1})$ using the Schur decomposition of the covariance matrix.  

This density is Gaussian with pdf $\mathcal{N}(\mu,\sigma^2)$. 
Note that in order to compute this pdf, we need to draw the samples ${c_{m-1}, \dots,c_{1}}$ which correspond to the embedding changes performed on the $m-1$ previous DCT coefficients. 
This can be done by sampling over the Gaussian distribution until the sample belong into the interval corresponding to the right embedding change. We can then compute the pfm by again integration over the 3 intervals $]-\infty,-q_{i,j}/2]$, $]-q_{i,j}/2,q_{i,j}/2]$, $]q_{i,j}/2,+\infty]$ for $-1, 0, +1$.

Once the pmf is computed, either we sample from it, or we convert the probabilities to costs using the relation $c_{i,j}^{k}=\log(p_{i,j}^{0}/p_{i,j}^{k})$, and use a STC.

  %Talk about the need of sampling in the continuous domain here.

\section{Results\label{sec:results}}
  \subsection{Database development\label{subsec:database}}
      Because BOSSBase has been built from differents cameras, the full-frame sensors files have differents sizes (from CR2 of size $2602\times3906$px, to DNG of size $3472\times5216$px, NEF of size $2014\times3039$px, and PEF files of size $3124\times4688$px), thus to be able to have the same down-sampling factor for each image it is important to find the minimum length or width dimension for all the images. 
      As a result, for each image we performed a centered crop of width and height equal to $l_{min}=2014$, and then we developed the image using bi-linear demosaicking, luminance averaging, bilinear downscaling and JPEG compression to build our BOSSBase-SD (same dimensions) for the given quality factors $QF95$. 
      Note that except for the crop operation and the demosaicking and down-sampling kernels, this database is very similar to the BOSSBase database.

  \subsection{Benchmark setup}
    %We evaluate the empirical security of our scheme in the JPEG domain from cropped and downsized images using the same downsizing ratio (see Section ~\ref{sec:Analysis}) producing a total of 10000 $512\times 512$ images from BOSSBase RAW to build the covers subset.

    The empirical security is evaluated as the minimal total classification error probability under equal priors, $P_{\mathrm{E}}=\min_{P_{\mathrm{FA}}}\frac{1}{2}(P_{\mathrm{FA}}+P_{\mathrm{MD}})$, with $P_{FA}$ and $P_{MD}$ standing for the false-alarm and missed detection rates. The JPEG images are steganalyzed with the DCTR feature set~\cite{holub2015low} and  the low-complexity linear classifier~\cite{cogranne2015ensemble}.

\begin{figure}[h]
      \begin{centering}
      \subfloat[]{\begin{centering}
      \includegraphics[width=0.45\columnwidth]{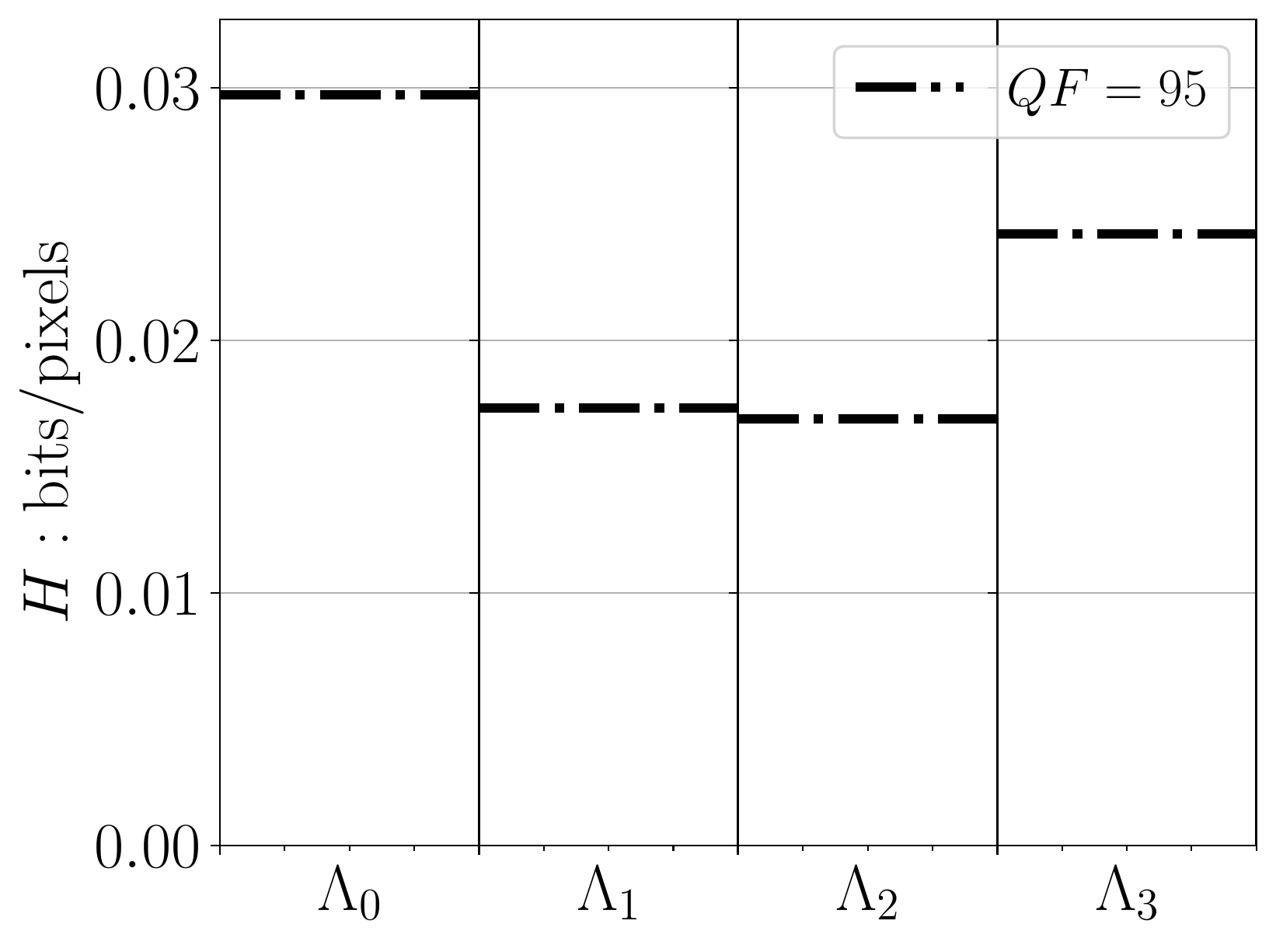}
      \par\end{centering}

      }\subfloat[]{\begin{centering}
      \includegraphics[width=0.45\columnwidth]{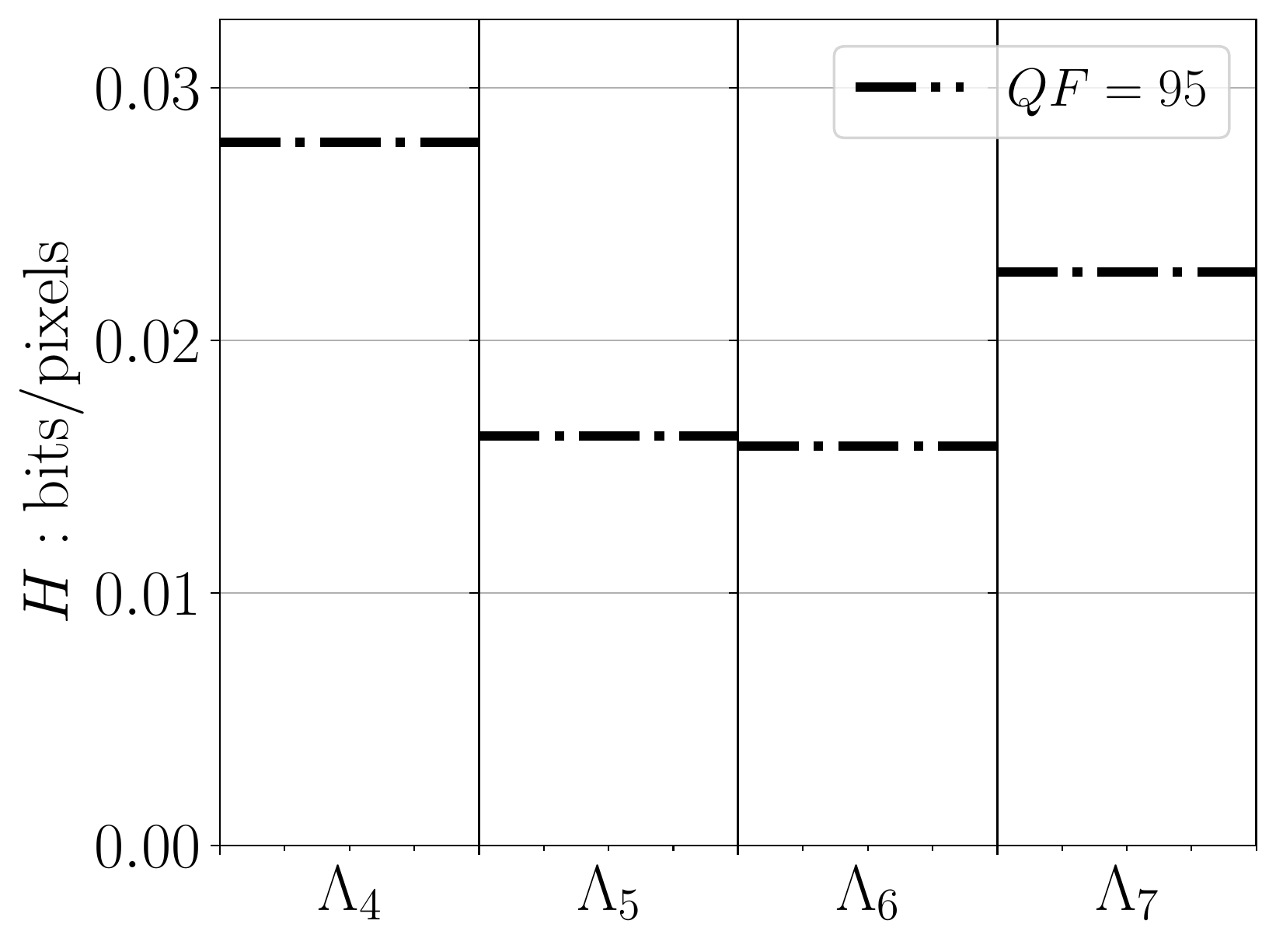}
      \par\end{centering}

      }
      \par\end{centering}
      \caption{Average entropy in bits/coefficients over the differents lattices : (a) intras-blocks, (b) inter-blocks.\label{fig:H_intra_inter}}
    \end{figure}

    The following embedding schemes are compared:

    \begin{itemize}
      \item {\it J-UNIWARD-Synchronized} and {\it UERD-Synchronized}: where the embedding pipeline used the estimated correlations matrix $\hat{\mathbf{R}}$ to build the dedicated covariances matrix $\hat{\mathbf{\Sigma}}$ using the embedding probabilities provided respectively by J-UNIWARD and UERD to perform the sampling. All theses step are explained at Figure ~\ref{fig:overview}.

      \item {\it J-UNIWARD} and {\it UERD}: Because the synchronized version of theses algorithm use random variables conditioning the achievable binary entropy is slightly attenuated as can be seen at Figure ~\ref{fig:H_DROP_95}, more details about how conditionning influence average entropy over each lattice can be observed at Figure ~\ref{fig:H_intra_inter}. 
      Therefore, in order to make an honest comparison we have compared J-UNIWARD and UERD to its synchronized version using the entropy obtained after conditioning.
    \end{itemize}

  \subsection{Comparison with UERD and J-UNIWARD}
    As mentioned above, our approach uses conditioned random variable sampling compared to previous realizations: the attainable binary entropy would be thus reduced as compared to the reachable entropy considering independant changesas it coule be observed on Figure where the conditioning induces a slight drop of the upper bound over the reachable entropy (less then $10^{-1}$ bits per coefficient). 

    \begin{figure}[H]
      \begin{centering}
      \includegraphics[width=0.8\columnwidth]{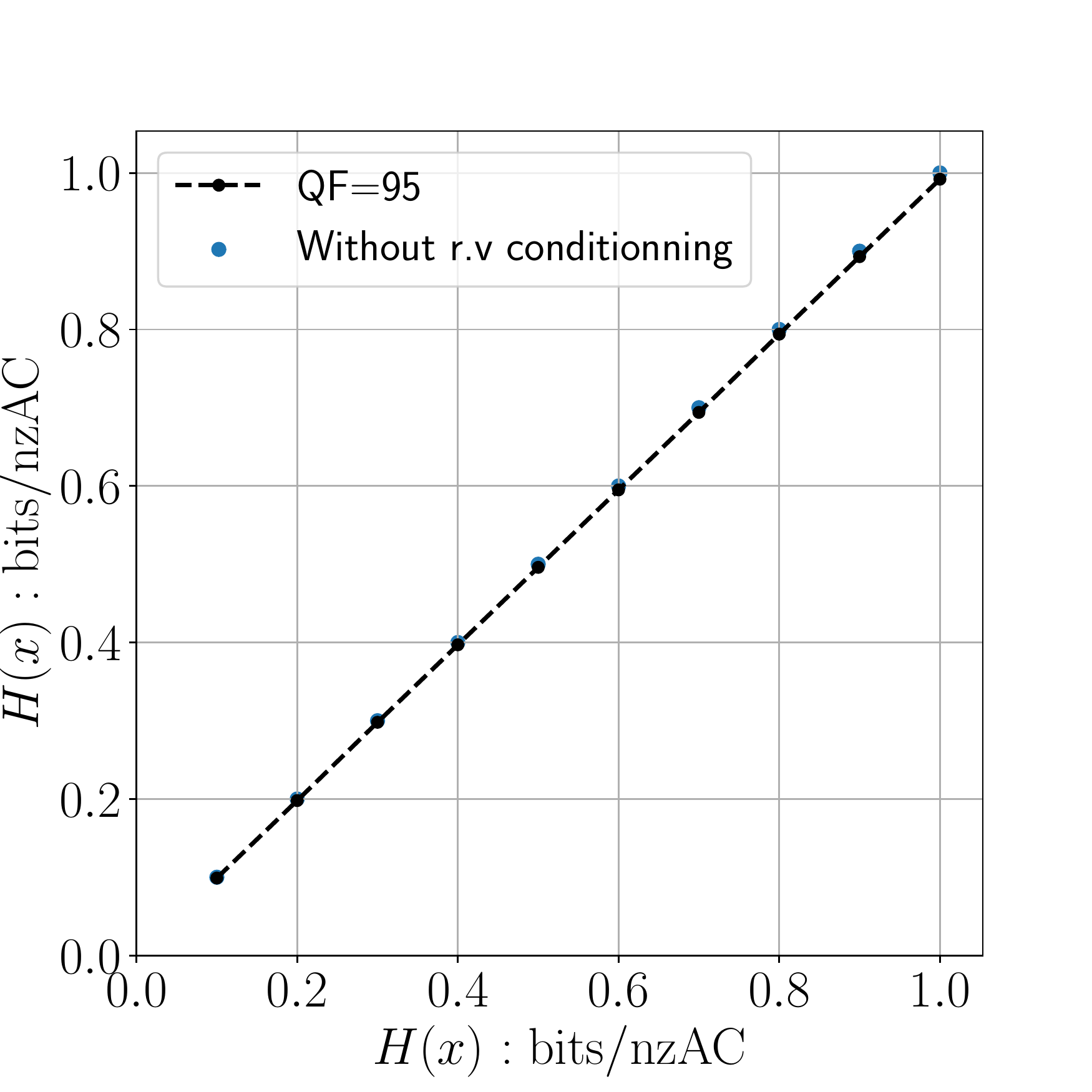}
      \par
      \end{centering}
      \caption{Average drop of entropy due to conditiong for $QF95$.\label{fig:H_DROP_95}}
    \end{figure}

\begin{figure}[h]
      \begin{centering}
      \includegraphics[width=0.8\columnwidth]{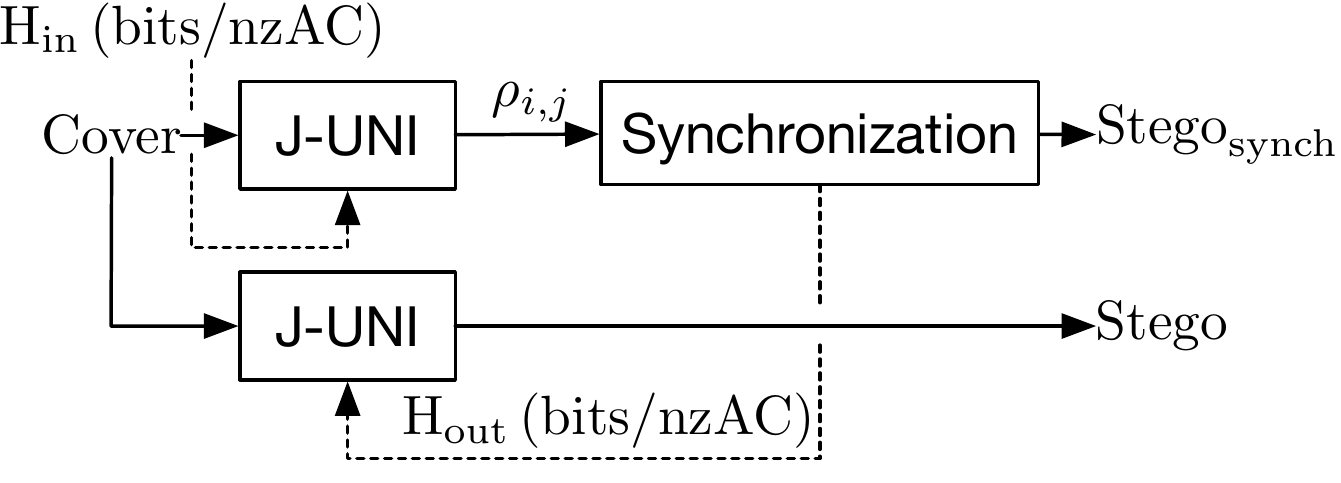}
      \par
      \end{centering}
      \caption{Embedding setup to ensure that the stego image carry the payload in the case of a J-UNIWARD embedding.\label{fig:scheme_JUNI_embedding_setup}}
    \end{figure}

    Thus, in order to perform a fair comparison between UERD, J-UNIWARD and their respectful synchronized version, for a given payload inputed $\mathrm{H_{in}\left(bits/nzAC\right)}$ the first step is to extract the costs from UERD or J-UNIWARD, compute the associated PMFs for each DCT coefficient, perform the synchronization and obtain the stego and compute the entropy $\mathrm{H_{out}\left(bits/nzAC\right)}$ achievable for the new PMFs for each DCT coefficient. Because we use r.v conditionning we have $\mathrm{H_{in}} > \mathrm{H_{out}}$ as we can observe at Figure ~\ref{fig:H_DROP_95}. To ensure that the $\mathrm{{Stego}_{synch}}$ and the non-synchronized $\mathrm{Stego}$ carry the same amount of information, $\mathrm{Stego}$ is embedded with the payload $\mathrm{H_{out}\left(bits/nzAC\right)}$ as depicted in Figure ~\ref{fig:scheme_JUNI_embedding_setup}.

    This operation is performed over the whole imagebase for $\mathrm{H_{in}\left(bits/nzAC\right)} \in \left\{ 0.1,\,0.2,\ldots,\,1.0\right\}$.

\begin{figure}[H]
      \begin{centering}
        \subfloat[$QF75$]{
          \begin{centering}
          \includegraphics[width=0.45\columnwidth]{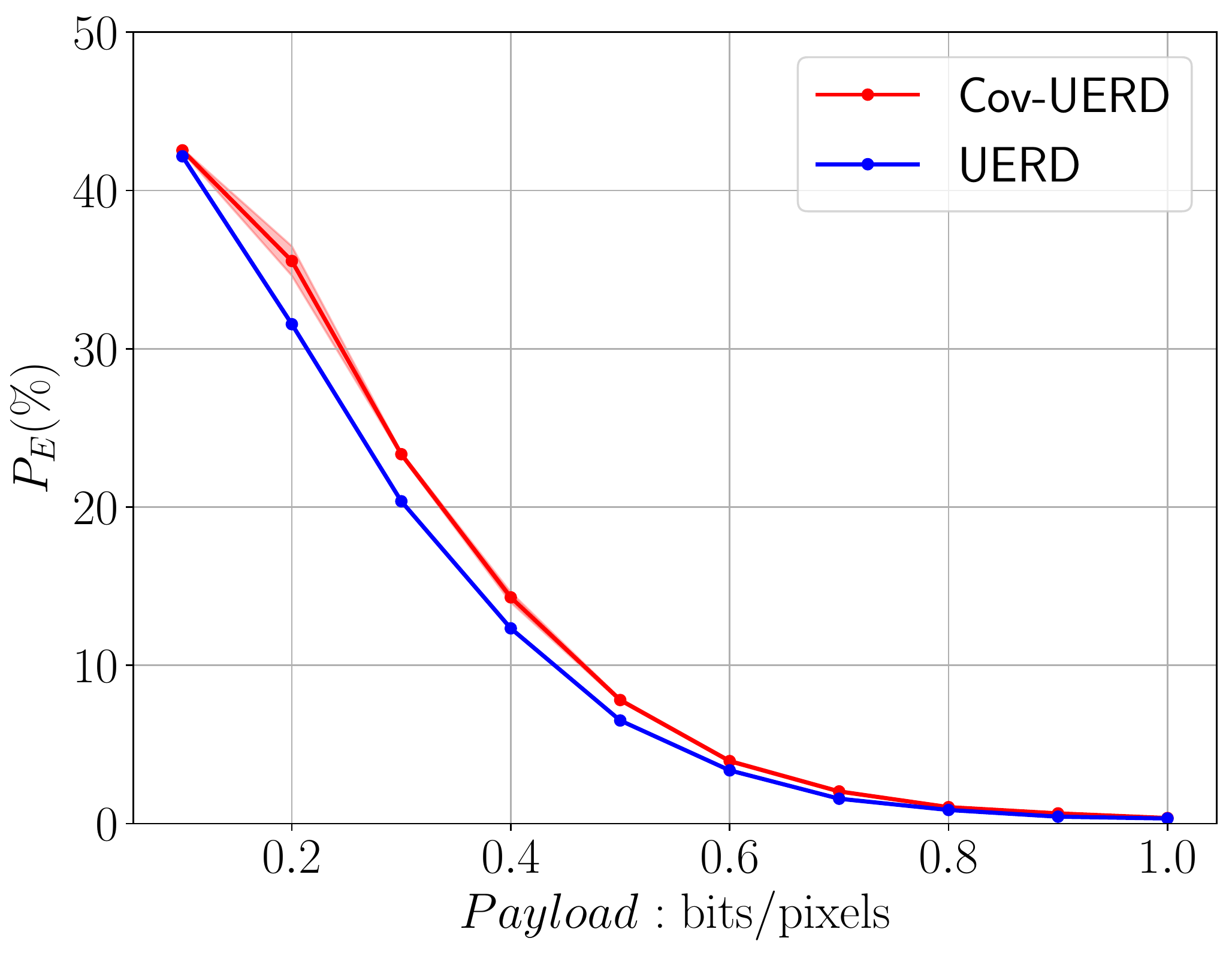}
          \par
          \end{centering}
        }
        \subfloat[$QF85$]{
          \begin{centering}
          \includegraphics[width=0.45\columnwidth]{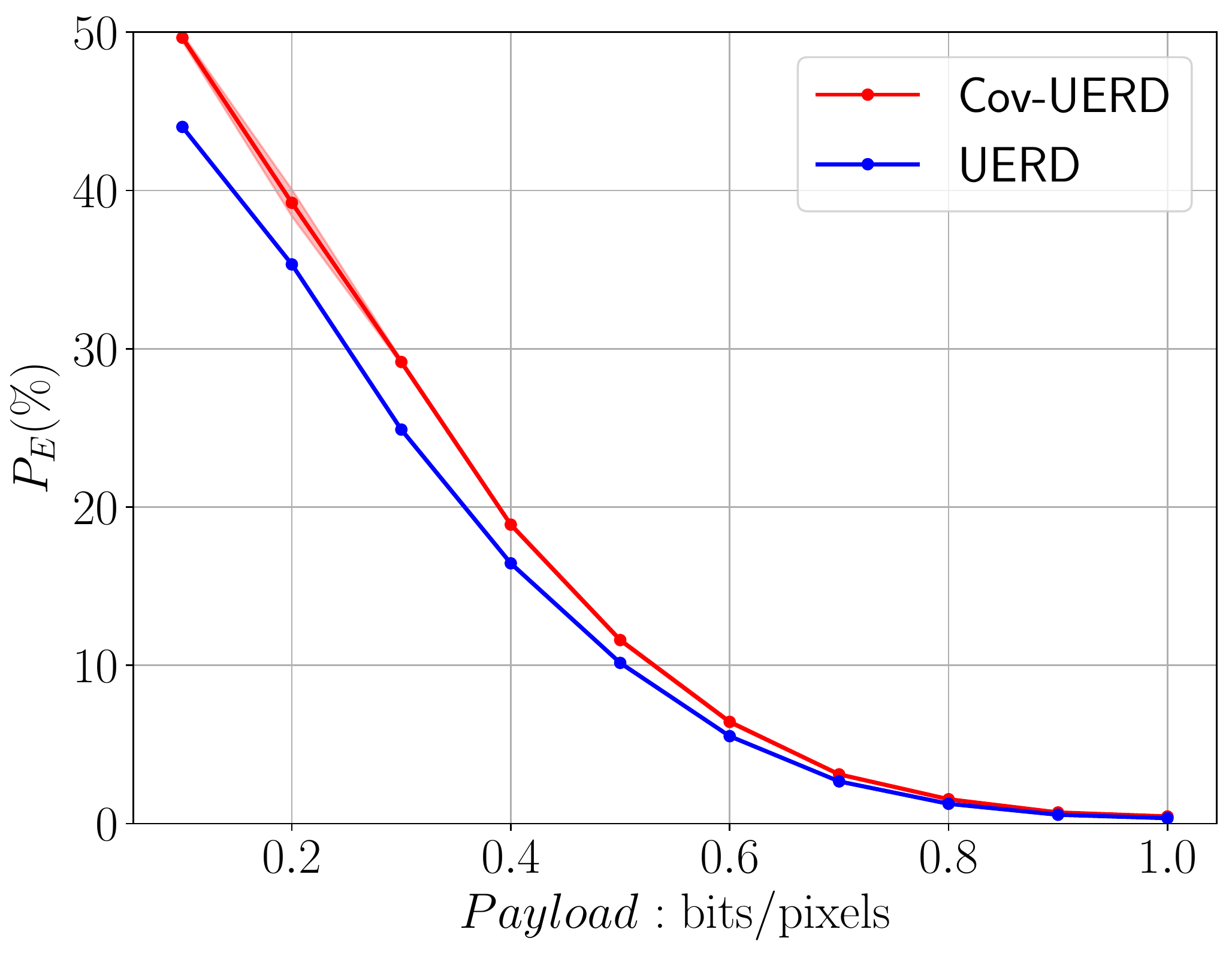}
          \par
          \end{centering}
        } \quad
        \subfloat[$QF95$]{
          \begin{centering}
          \includegraphics[width=0.45\columnwidth]{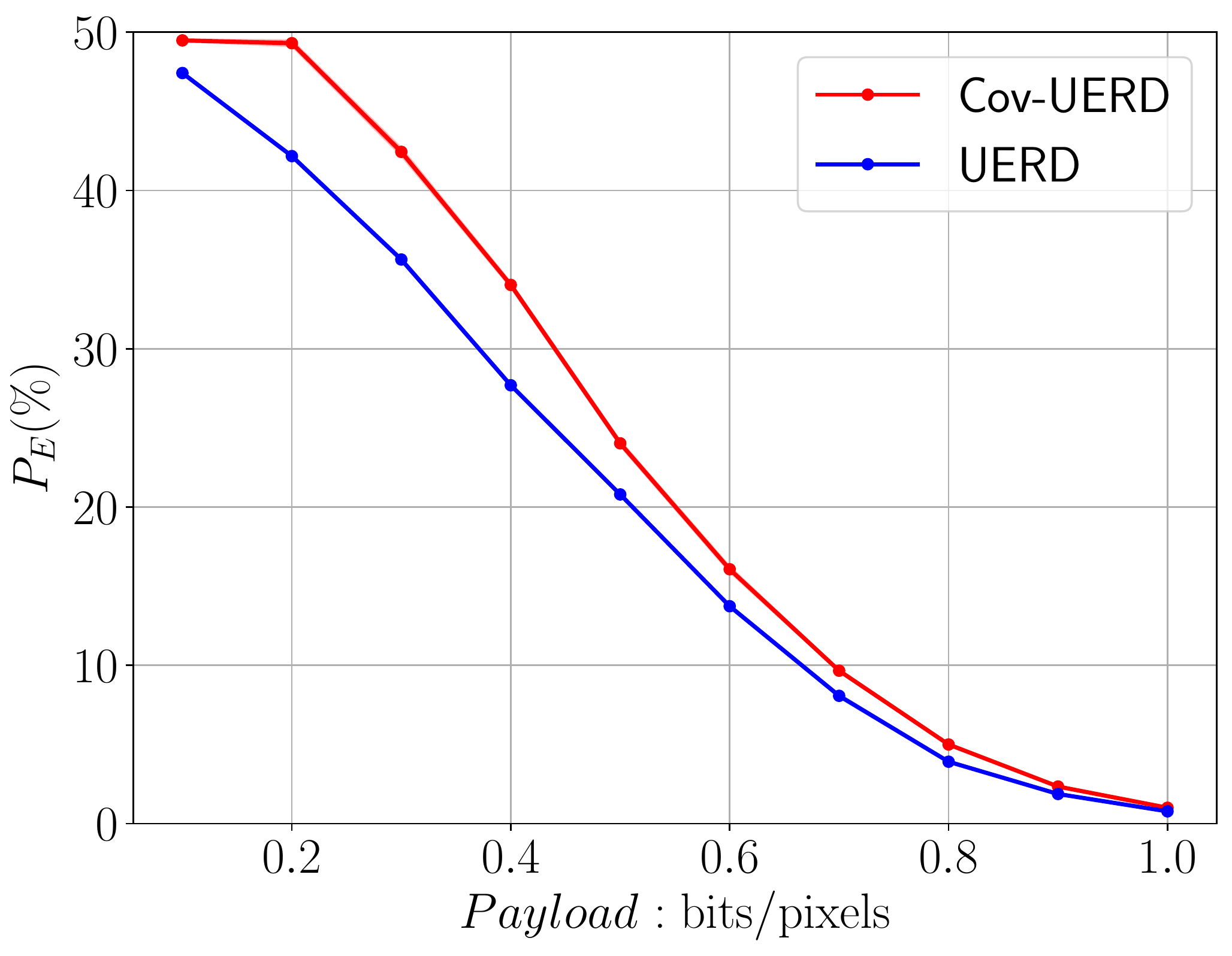}
          \par
          \end{centering}
        }
        \subfloat[$QF100$]{
          \begin{centering}
          \includegraphics[width=0.45\columnwidth]{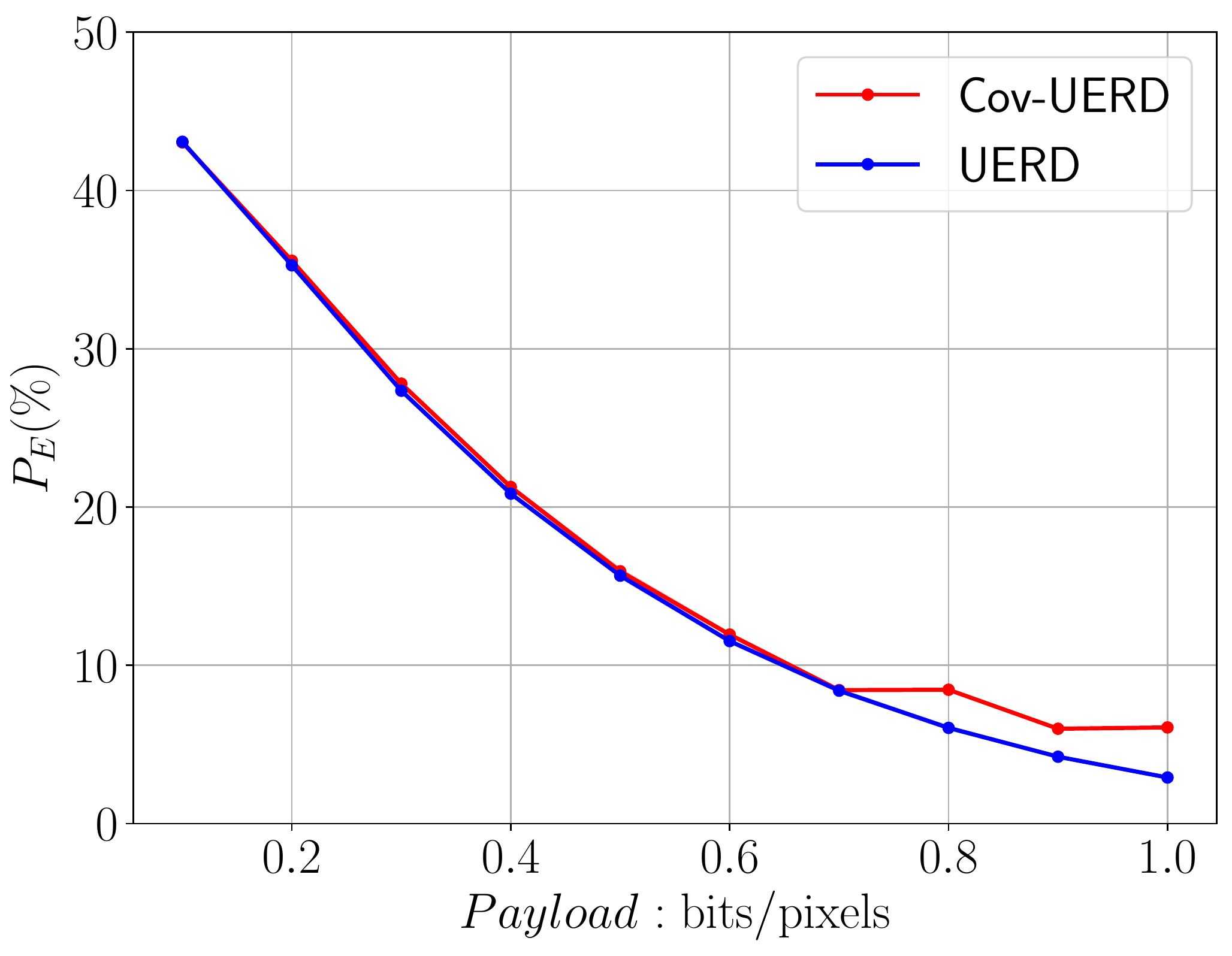}
          \par
          \end{centering}
        }
      \par\end{centering}
      \caption{UERD and its synchronized version $QF\in\{75, 95, 85, 100\}$ for respectively (a), (b), (c) and (d).\label{fig:PE_UERD_75_85_95_100}}
    \end{figure}

\begin{figure}[H]
      \begin{centering}
        \subfloat[$QF75$]{
          \begin{centering}
          \includegraphics[width=0.45\columnwidth]{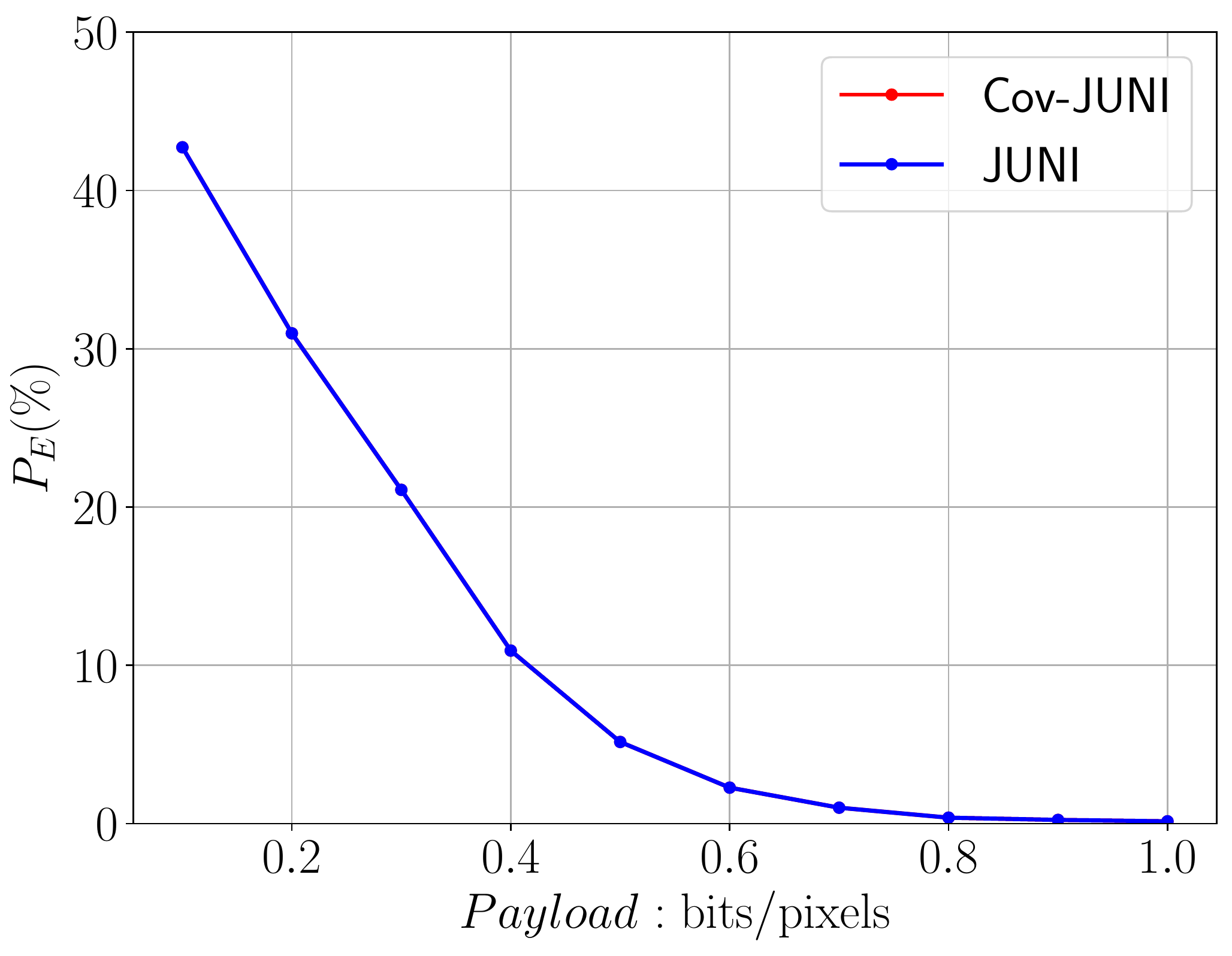}
          \par
          \end{centering}
        }
        \subfloat[$QF85$]{
          \begin{centering}
          \includegraphics[width=0.45\columnwidth]{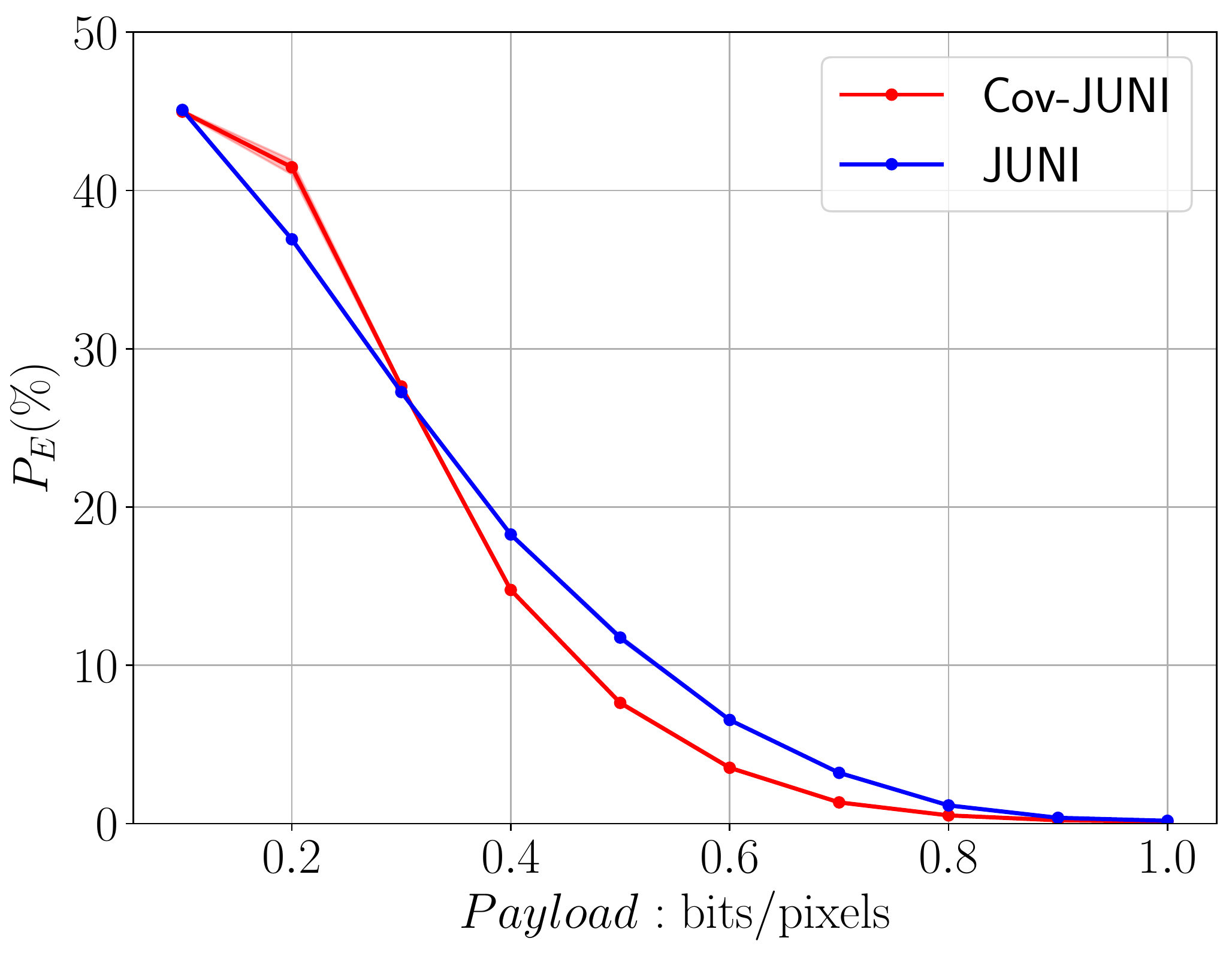}
          \par
          \end{centering}
        } \quad
        \subfloat[$QF95$]{
          \begin{centering}
          \includegraphics[width=0.45\columnwidth]{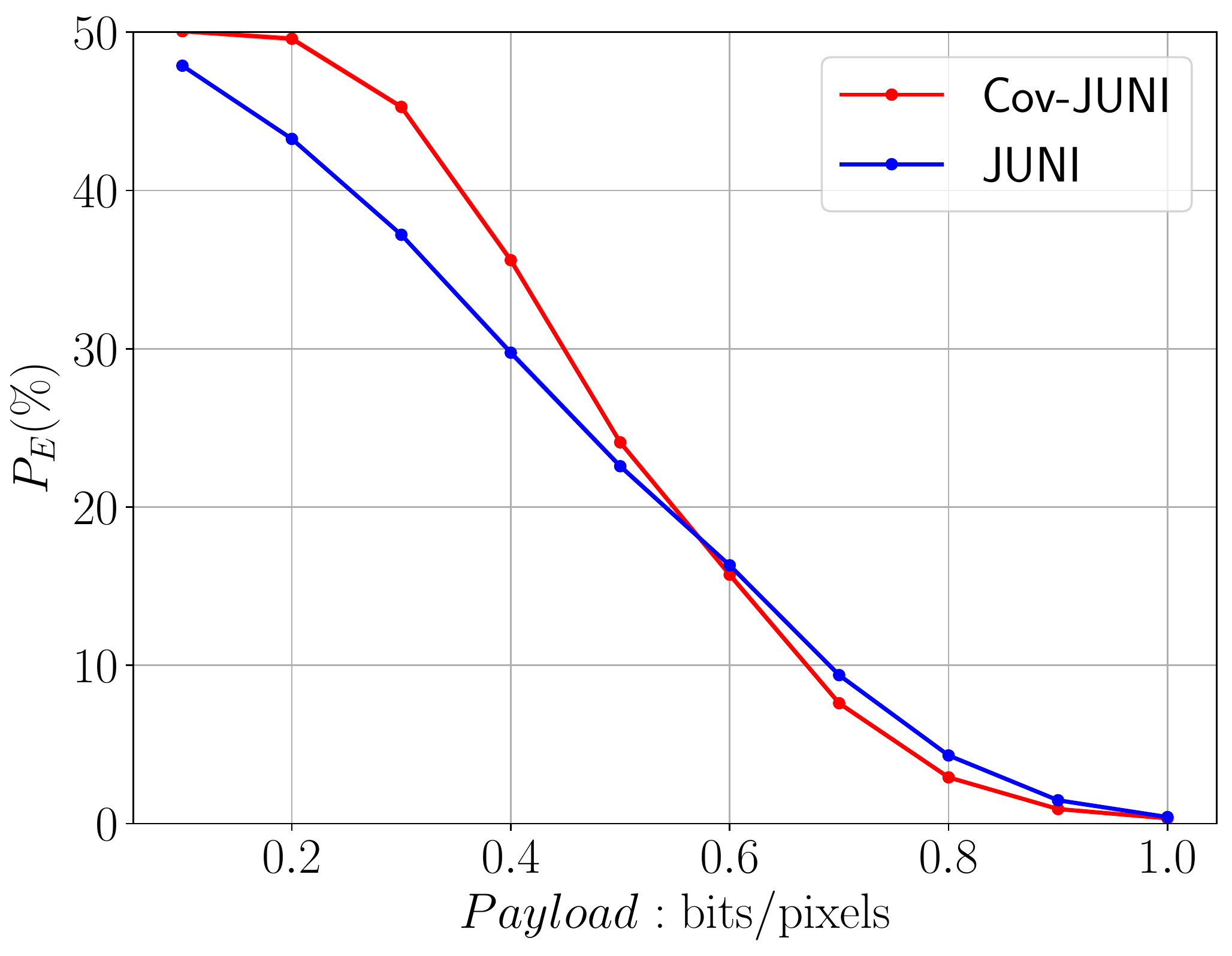}
          \par
          \end{centering}
        }
        \subfloat[$QF100$]{
          \begin{centering}
          \includegraphics[width=0.45\columnwidth]{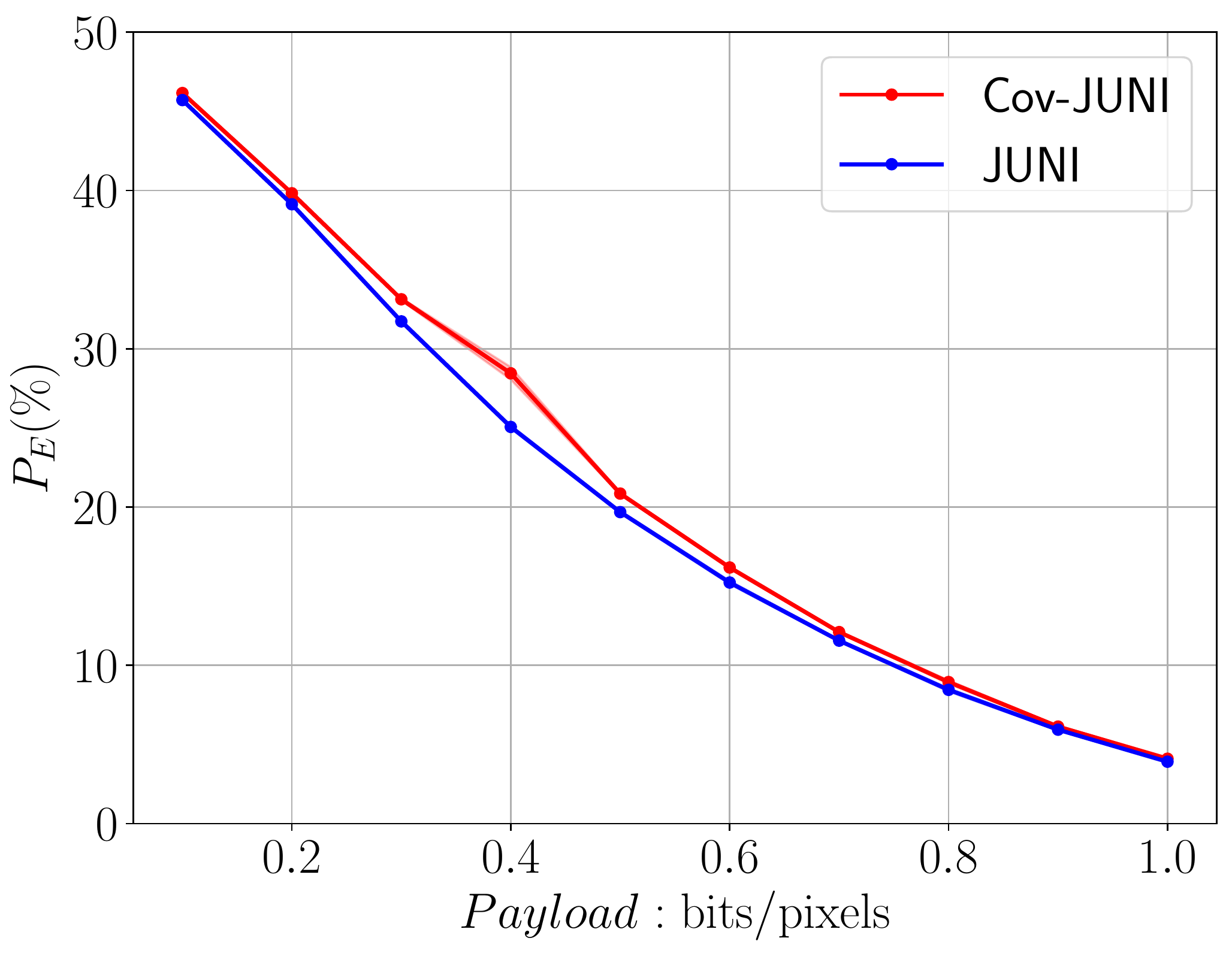}
          \par
          \end{centering}
        }
      \par\end{centering}
      \caption{J-UNIWARD and its synchronized version for $QF\in\{75, 95, 85, 100\}$ for respectively (a), (b), (c) and (d).\label{fig:PE_JUNI_75_85_95_100}}
    \end{figure}

    For UERD at $QF95$ and a payload of between $0.1$ and $0.4$ bpnzAC, this leads to an increase of the empirical security of a little more than $5\%$ as outlined in the table ~\ref{tab:tab_PE_payload_0_3}. The increase in empirical security can be observed over all points obtained from UERD between $0.1$ and $1.0$ bpnzAC for $QF\in\{75, 95, 85, 100\}$, achieving up to a $+7\%$ gain of $P_E$ for QF95 and an insertion rate set at $0.28$ for both embedding algorithms.
    However, our approach does not provide the same stability of empirical security for J-UNIWARD, see Figure ~\ref{fig:PE_UERD_75_85_95_100} and ~\ref{fig:PE_JUNI_75_85_95_100}.

    \begin{table}[t]
      \begin{adjustbox}{width=\columnwidth, center}
        \begin{centering}
        \begin{tabular}{|c|c|c|c|c|}
        \hline 
        $P_{E}$ (\%) / & Cov-UERD & UERD & Cov-JUNI & JUNI\tabularnewline
        JPEG QF &  &  & & \tabularnewline
        \hline 
        \hline 
        % 100 & x & x & x & x\tabularnewline
        % \hline 
        75 & 23.341 $\pm$ 0.116 & 20.368 $\pm$ 0.08 & 21.089 $\pm$ 0.104 & 21.606 $\pm$ 0.059\tabularnewline
        \hline
        85 & 29.167 $\pm$ 0.145 & 24.896 $\pm$ 0.11 & 27.62 $\pm$ 0.136 & 27.269 $\pm$ 0.109\tabularnewline
        \hline
        95 & 42.442 $\pm$ 0.242 & 35.64 $\pm$ 0.11 & 45.282 $\pm$ 0.113 & 37.205 $\pm$ 0.045\tabularnewline
        \hline
        100 & 27.797 $\pm$ 0.094 & 27.351 $\pm$ 0.069 & 33.129 $\pm$ 0.08 & 31.733 $\pm$ 0.095\tabularnewline
        \hline 
        \end{tabular}
        \par\end{centering}
      \end{adjustbox}
      \medskip{}

      \caption{Average empirical security ($P_{\mathrm{E}}$ in \%) over 10 runs for different quality factors and embedding strategies on BOSSBase SD with bilinear demoisaicking, and downscaling but the same payload of $0.28$ bpnzac. DCTR features combined with regularized linear classifier are used for steganalysis.\label{tab:tab_PE_payload_0_3}}
    \end{table}

  \subsection{Effects of synchronization}

\begin{figure}[h]
      \begin{centering}
      \includegraphics[width=0.8\columnwidth]{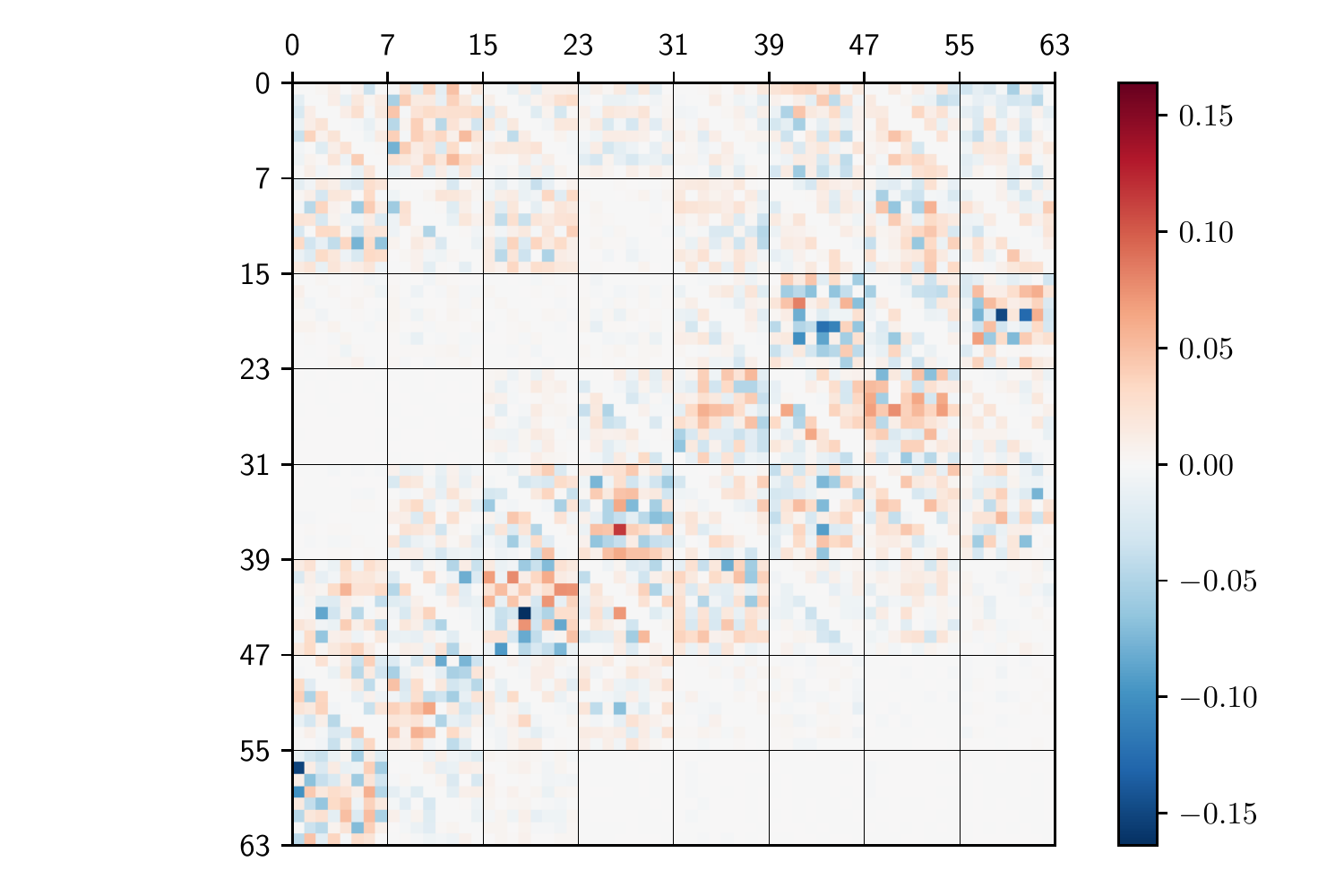}
      \par
      \end{centering}
      \caption{Difference of probabilities map to sample a $+1$ between UERD and Cov-UERD for a sample image (cropped to a $64 \times 64$ array) of $QF100$ embedded at a $0.48$ bpnzAC rate. Identical embedding changes for the two schemes have been performed on coefficients belonging to lattice $\Lambda_0$.\label{fig:P_map_Diff_p1}}
    \end{figure}

\begin{figure}[h]
      \begin{centering}
      \includegraphics[width=0.9\columnwidth]{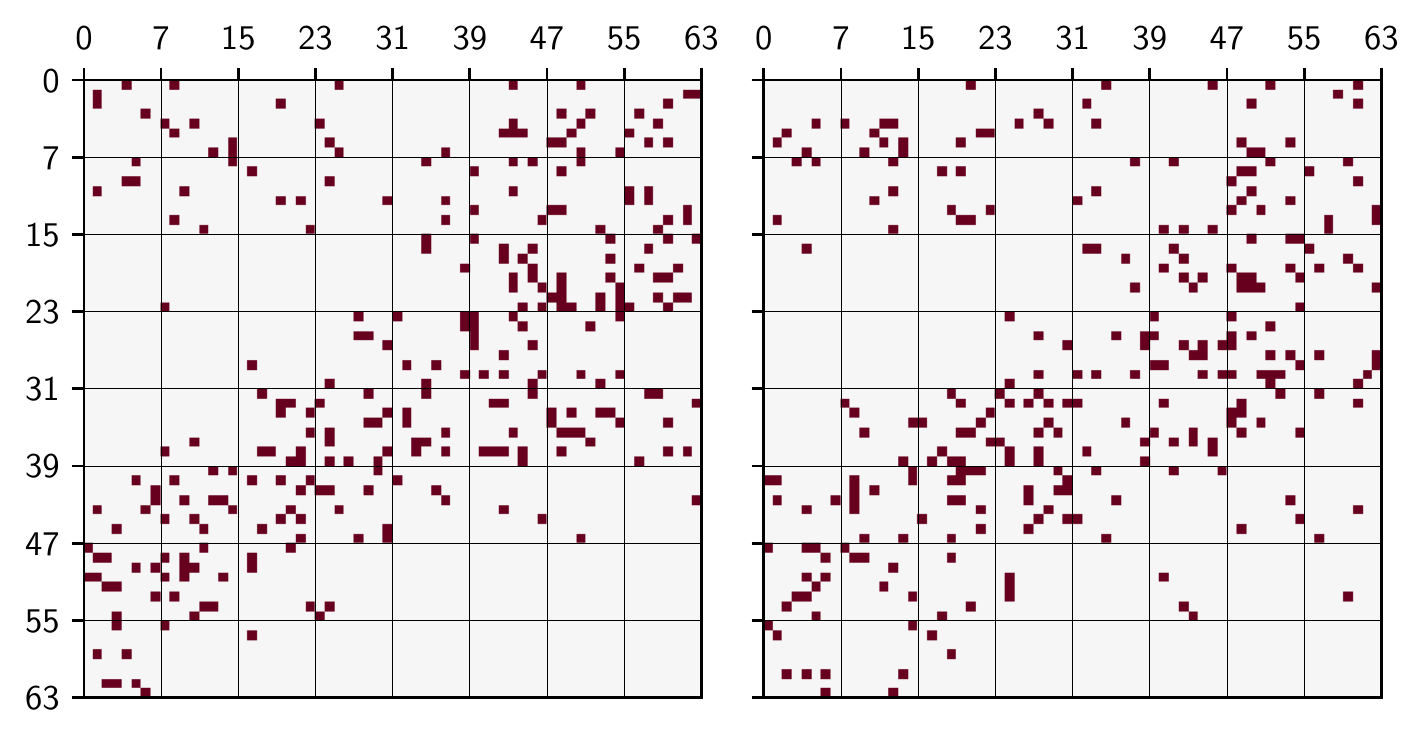}
      \par
      \end{centering}
      \caption{$+1$ modifications performed for respectively UERD (left) and its synchronized version (right).\label{fig:P1_embedding_UERD_and_SYNCH}}
    \end{figure}

    The synchronization w.r.t. previous embedding changes naturally induces fluctuations in the final embedding probabilities. 
    One can observe on figure ~\ref{fig:P_map_Diff_p1} that the embedding probabilities of the coefficients belonging to $\mathbf{\Lambda_0}$ remain unchanged, but that other probabilities can undergo an important bias, going up to $\pm 0.15$.
    However, using covariance matrices as an oracle of correlations allows us to encourage changes in modes that would not have been changed at all or not in the same polarity as depicted on Figure ~\ref{fig:P1_embedding_UERD_and_SYNCH} where for an cropped sample image we plot the $+1$ modifications which were carried out in the case of an embedding using the synchronized version of UERD and UERD (without synchronization) featured with the entropy reached after conditioning. 

  \subsection{Complexity}
    This embedding algorithm is computationally expensive because the complexity of computing the conditional distribution increases with the complexity of the Cholesky decomposition of the covariance matrix, i.e., as $\mathcal{O}(n^{3})$ where $n = \mathrm{Card}\left(\mathrm{idx}(m)\right) + 1$, depending of the mode $m$ and to which lattice its belong : $n=1$ for $m \in \mathbf{\Lambda}_{0}$, $n=3$ for $m \in \mathbf{\Lambda}_{3}$ and $n=39$ for $m \in \mathbf{\Lambda}_{7}$. On a 1.6 GHz Intel Core i5, our python implementation of simulated embedding on a $512\times512$ image is performed in 1min 46s while an UERD simulated embedding take 2 seconds.

  \section{Conclusions and perspectives}
    We have proposed a synchronization mechanism for JPEG steganography that can be used for classical additive cost-based embedding schemes. The synchronization is done by incorporating correlations estimated between DCT coefficients after the development of the image from RAW to JPEG. Our encouraging results shows that this rational enables to increase the practical undetectability by around $7\%$ for an embedding rate of 0.28 bpnzac. In our future works, we plan to compare our results with \cite{zhang2016decomposing} using the same data-base, and use appropriate quantization steps for $QF \in \{75, 85, 95\}$ ($q=1$ was used).

  \section{Appendix}

    % \begin{table*}[h]
    %   \begin{adjustbox}{width=1.8\columnwidth, center}
    %     \begin{centering}
    %
%\begin{tabular}{|c|c|c|c|c|c|c|c|c|c|c|c|c|c|c|c|c|}
    %       \hline Mode / &\multirow{2}{*}{$(0, 0)$} & \multirow{2}{*}{$(1, 1)$} & \multirow{2}{*}{$(2, 2)$} & \multirow{2}{*}{$(3, 3)$} & \multirow{2}{*}{$(4, 4)$} & \multirow{2}{*}{$(5, 5)$} & \multirow{2}{*}{$(6, 6)$} & \multirow{2}{*}{$(7, 7)$} & \multirow{2}{*}{$(0, 1)$} & \multirow{2}{*}{$(1, 2)$} & \multirow{2}{*}{$(2, 3)$} & \multirow{2}{*}{$(3, 4)$} & \multirow{2}{*}{$(4, 5)$} & \multirow{2}{*}{$(5, 6)$} & \multirow{2}{*}{$(6, 7)$} & \multirow{2}{*}{$(7, 0)$}
    %       \tabularnewline Bloc & & & & & & & & & & & & & & & & \tabularnewline
    %       \hline
    %               \multirow{1}{*}{0} & (0, 0) & (1, 1) & (2, 2) & (3, 3) & (4, 4) & (5, 5) & (6, 6) & (7, 7) & (0, 1) & (1, 2) & (2, 3) & (3, 4) & (4, 5) & (5, 6) & (6, 7) & (7, 0)\tabularnewline
    %       \cline{2-17} \cline{3-17} \cline{4-17} \cline{5-17} \cline{6-17} \cline{7-17} \cline{8-17} \cline{9-17}
    %       \cline{9-17} \cline{10-17} \cline{5-17} \cline{11-17} \cline{12-17} \cline{13-17} \cline{14-17} \cline{15-17}
    %       \cline{16-17} \cline{17-17}
    %       \hline 
    %       \end{tabular}
    %     \par\end{centering}
    %   \end{adjustbox}
    %   \medskip{}

    %   \caption{Correlated mode block n°1, 2, 3 or 4 for each mode of $\mathbf{\Lambda_{0}}$.}\label{tab_L0}
    % \end{table*}

    \begin{table*}[h]
      \begin{adjustbox}{width=1.8\columnwidth, center}
      \begin{centering}
      \begin{tabular}{|c|c|c|c|c|c|c|c|c|c|c|c|c|c|c|c|c|}
      \hline Mode / &\multirow{2}{*}{$(0, 2)$} & \multirow{2}{*}{$(1, 3)$} & \multirow{2}{*}{$(2, 4)$} & \multirow{2}{*}{$(3, 5)$} & \multirow{2}{*}{$(4, 6)$} & \multirow{2}{*}{$(5, 7)$} & \multirow{2}{*}{$(7, 1)$} & \multirow{2}{*}{$(6, 0)$} & \multirow{2}{*}{$(0, 3)$} & \multirow{2}{*}{$(1, 4)$} & \multirow{2}{*}{$(2, 5)$} & \multirow{2}{*}{$(3, 6)$} & \multirow{2}{*}{$(4, 7)$} & \multirow{2}{*}{$(7, 2)$} & \multirow{2}{*}{$(6, 1)$} & \multirow{2}{*}{$(5, 0)$}
      \tabularnewline Block & & & & & & & & & & & & & & & & \tabularnewline
      \hline
          \multirow{3}{*}{0} & (0, 0) & (1, 1) & (2, 2) & (3, 3) & (4, 4) & (5, 5) & (1, 1) & (0, 0) & (0, 1) & (1, 2) & (2, 3) & (3, 4) & (4, 5) & (1, 2) & (0, 1) & (5, 6)\tabularnewline
          \cline{2-17} \cline{3-17} \cline{4-17} \cline{5-17} \cline{6-17} \cline{7-17} \cline{8-17} \cline{9-17}
          \cline{9-17} \cline{10-17} \cline{5-17} \cline{11-17} \cline{12-17} \cline{13-17} \cline{14-17} \cline{15-17}
          \cline{16-17} \cline{17-17}
          & (2, 2) & (3, 3) & (4, 4) & (5, 5) & (6, 6) & (7, 7) & (7, 7) & (6, 6) & (2, 3) & (3, 4) & (4, 5) & (5, 6) & (6, 7) & (7, 0) & (6, 7) & (7, 0)\tabularnewline
          \cline{2-17} \cline{3-17} \cline{4-17} \cline{5-17} \cline{6-17} \cline{7-17} \cline{8-17} \cline{9-17}
          \cline{9-17} \cline{10-17} \cline{5-17} \cline{11-17} \cline{12-17} \cline{13-17} \cline{14-17} \cline{15-17}
          \cline{16-17} \cline{17-17}
          \cline{2-17} \cline{3-17} \cline{4-17} \cline{5-17} \cline{6-17} \cline{7-17} \cline{8-17} \cline{9-17}
          \cline{9-17} \cline{10-17} \cline{5-17} \cline{11-17} \cline{12-17} \cline{13-17} \cline{14-17} \cline{15-17}
          \cline{16-17} \cline{17-17}
          \hline 
      \end{tabular}
      \par\end{centering}
      \end{adjustbox}
      \medskip{}

      \caption{Correlated modes for each mode of $\mathbf{\Lambda_{1}}$ w.r.t. coefficients belonging to the previous lattice.}\label{tab_L1}
    \end{table*}

    \begin{table*}[h]
      \begin{adjustbox}{width=1.8\columnwidth, center}
      \begin{centering}
      \begin{tabular}{|c|c|c|c|c|c|c|c|c|c|c|c|c|c|c|c|c|}
      \hline Mode / &\multirow{2}{*}{$(0, 4)$} & \multirow{2}{*}{$(1, 5)$} & \multirow{2}{*}{$(2, 6)$} & \multirow{2}{*}{$(3, 7)$} & \multirow{2}{*}{$(7, 3)$} & \multirow{2}{*}{$(6, 2)$} & \multirow{2}{*}{$(5, 1)$} & \multirow{2}{*}{$(4, 0)$} & \multirow{2}{*}{$(0, 5)$} & \multirow{2}{*}{$(1, 6)$} & \multirow{2}{*}{$(2, 7)$} & \multirow{2}{*}{$(7, 4)$} & \multirow{2}{*}{$(6, 3)$} & \multirow{2}{*}{$(5, 2)$} & \multirow{2}{*}{$(4, 1)$} & \multirow{2}{*}{$(3, 0)$}
      \tabularnewline Block & & & & & & & & & & & & & & & & \tabularnewline
      \hline
          \multirow{5}{*}{0} & (0, 0) & (1, 1) & (2, 2) & (3, 3) & (3, 3) & (2, 2) & (1, 1) & (0, 0) & (0, 1) & (1, 2) & (2, 3) & (3, 4) & (2, 3) & (1, 2) & (0, 1) & (3, 4)\tabularnewline
          \cline{2-17} \cline{3-17} \cline{4-17} \cline{5-17} \cline{6-17} \cline{7-17} \cline{8-17} \cline{9-17}
          \cline{9-17} \cline{10-17} \cline{5-17} \cline{11-17} \cline{12-17} \cline{13-17} \cline{14-17} \cline{15-17}
          \cline{16-17} \cline{17-17}
          & (4, 4) & (5, 5) & (6, 6) & (7, 7) & (7, 7) & (6, 6) & (5, 5) & (4, 4) & (4, 5) & (5, 6) & (6, 7) & (7, 0) & (6, 7) & (5, 6) & (4, 5) & (7, 0)\tabularnewline
          \cline{2-17} \cline{3-17} \cline{4-17} \cline{5-17} \cline{6-17} \cline{7-17} \cline{8-17} \cline{9-17}
          \cline{9-17} \cline{10-17} \cline{5-17} \cline{11-17} \cline{12-17} \cline{13-17} \cline{14-17} \cline{15-17}
          \cline{16-17} \cline{17-17}
          & (0, 2) & (1, 3) & (2, 4) & (3, 5) & (1, 3) & (0, 2) & (5, 7) & (4, 6) & (0, 3) & (1, 4) & (2, 5) & (1, 4) & (0, 3) & (7, 2) & (4, 7) & (3, 6)\tabularnewline
\hline 
      \end{tabular}
      \par\end{centering}
      \end{adjustbox}
      \medskip{}

      \caption{Correlated modes for each mode of $\mathbf{\Lambda_{2}}$  w.r.t. coefficients belonging to the previous lattices.}\label{tab_L2}
    \end{table*}

    \begin{table*}[h]
      \begin{adjustbox}{width=1.8\columnwidth, center}
      \begin{centering}
      \begin{tabular}{|c|c|c|c|c|c|c|c|c|c|c|c|c|c|c|c|c|}
      \hline Mode / &\multirow{2}{*}{$(0, 6)$} & \multirow{2}{*}{$(1, 7)$} & \multirow{2}{*}{$(7, 5)$} & \multirow{2}{*}{$(6, 4)$} & \multirow{2}{*}{$(5, 3)$} & \multirow{2}{*}{$(4, 2)$} & \multirow{2}{*}{$(3, 1)$} & \multirow{2}{*}{$(2, 0)$} & \multirow{2}{*}{$(0, 7)$} & \multirow{2}{*}{$(7, 6)$} & \multirow{2}{*}{$(6, 5)$} & \multirow{2}{*}{$(5, 4)$} & \multirow{2}{*}{$(4, 3)$} & \multirow{2}{*}{$(3, 2)$} & \multirow{2}{*}{$(2, 1)$} & \multirow{2}{*}{$(1, 0)$}
      \tabularnewline Block & & & & & & & & & & & & & & & & \tabularnewline
      \hline
          \multirow{7}{*}{0} & (0, 0) & (1, 1) & (5, 5) & (4, 4) & (3, 3) & (2, 2) & (1, 1) & (0, 0) & (0, 1) & (5, 6) & (4, 5) & (3, 4) & (2, 3) & (1, 2) & (0, 1) & (1, 2)\tabularnewline
          \cline{2-17} \cline{3-17} \cline{4-17} \cline{5-17} \cline{6-17} \cline{7-17} \cline{8-17} \cline{9-17}
          \cline{9-17} \cline{10-17} \cline{5-17} \cline{11-17} \cline{12-17} \cline{13-17} \cline{14-17} \cline{15-17}
          \cline{16-17} \cline{17-17}
          & (6, 6) & (7, 7) & (7, 7) & (6, 6) & (5, 5) & (4, 4) & (3, 3) & (2, 2) & (6, 7) & (7, 0) & (6, 7) & (5, 6) & (4, 5) & (3, 4) & (2, 3) & (7, 0)\tabularnewline
          \cline{2-17} \cline{3-17} \cline{4-17} \cline{5-17} \cline{6-17} \cline{7-17} \cline{8-17} \cline{9-17}
          \cline{9-17} \cline{10-17} \cline{5-17} \cline{11-17} \cline{12-17} \cline{13-17} \cline{14-17} \cline{15-17}
          \cline{16-17} \cline{17-17}
          & (0, 2) & (1, 3) & (3, 5) & (2, 4) & (1, 3) & (0, 2) & (3, 5) & (2, 4) & (0, 3) & (3, 6) & (2, 5) & (1, 4) & (0, 3) & (3, 6) & (2, 5) & (1, 4)\tabularnewline
\hline 
      \end{tabular}
      \par\end{centering}
      \end{adjustbox}
      \medskip{}

      \caption{Correlated modes for each mode of $\mathbf{\Lambda_{3}}$  w.r.t. coefficients belonging to the previous lattices.}\label{tab_L3}
    \end{table*}

    \begin{table*}[h]
      \begin{adjustbox}{width=1.8\columnwidth, center}
      \begin{centering}
      \begin{tabular}{|c|c|c|c|c|c|c|c|c|c|c|c|c|c|c|c|c|}
      \hline Mode / &\multirow{2}{*}{$(0, 0)$} & \multirow{2}{*}{$(1, 1)$} & \multirow{2}{*}{$(2, 2)$} & \multirow{2}{*}{$(3, 3)$} & \multirow{2}{*}{$(4, 4)$} & \multirow{2}{*}{$(5, 5)$} & \multirow{2}{*}{$(6, 6)$} & \multirow{2}{*}{$(7, 7)$} & \multirow{2}{*}{$(0, 1)$} & \multirow{2}{*}{$(1, 2)$} & \multirow{2}{*}{$(2, 3)$} & \multirow{2}{*}{$(3, 4)$} & \multirow{2}{*}{$(4, 5)$} & \multirow{2}{*}{$(5, 6)$} & \multirow{2}{*}{$(6, 7)$} & \multirow{2}{*}{$(7, 0)$}
      \tabularnewline Block & & & & & & & & & & & & & & & & \tabularnewline
      \hline
          \multirow{8}{*}{1} & (0, 0) & (0, 1) & (0, 2) & (0, 3) & (0, 4) & (0, 5) & (0, 6) & (0, 7) & (0, 1) & (0, 2) & (0, 3) & (0, 4) & (0, 5) & (0, 6) & (0, 7) & (0, 0)\tabularnewline
          \cline{2-17} \cline{3-17} \cline{4-17} \cline{5-17} \cline{6-17} \cline{7-17} \cline{8-17} \cline{9-17}
          \cline{9-17} \cline{10-17} \cline{5-17} \cline{11-17} \cline{12-17} \cline{13-17} \cline{14-17} \cline{15-17}
          \cline{16-17} \cline{17-17}
          & (1, 0) & (1, 1) & (1, 2) & (1, 3) & (1, 4) & (1, 5) & (1, 6) & (1, 7) & (1, 1) & (1, 2) & (1, 3) & (1, 4) & (1, 5) & (1, 6) & (1, 7) & (1, 0)\tabularnewline
          \cline{2-17} \cline{3-17} \cline{4-17} \cline{5-17} \cline{6-17} \cline{7-17} \cline{8-17} \cline{9-17}
          \cline{9-17} \cline{10-17} \cline{5-17} \cline{11-17} \cline{12-17} \cline{13-17} \cline{14-17} \cline{15-17}
          \cline{16-17} \cline{17-17}
          & (2, 0) & (2, 1) & (2, 2) & (2, 3) & (2, 4) & (2, 5) & (2, 6) & (2, 7) & (2, 1) & (2, 2) & (2, 3) & (2, 4) & (2, 5) & (2, 6) & (2, 7) & (2, 0)\tabularnewline
\hline
      \end{tabular}
      \par\end{centering}
      \end{adjustbox}
      \medskip{}

      \caption{Correlated modes for each mode of $\mathbf{\Lambda_{4}}$ w.r.t. coefficients belonging to the previous lattices.}\label{tab_L4}
    \end{table*}

    \begin{table*}[h]
      \begin{adjustbox}{width=1.8\columnwidth, center}
      \begin{centering}
      \begin{tabular}{|c|c|c|c|c|c|c|c|c|c|c|c|c|c|c|c|c|}
      \hline Mode / &\multirow{2}{*}{$(0, 2)$} & \multirow{2}{*}{$(1, 3)$} & \multirow{2}{*}{$(2, 4)$} & \multirow{2}{*}{$(3, 5)$} & \multirow{2}{*}{$(4, 6)$} & \multirow{2}{*}{$(5, 7)$} & \multirow{2}{*}{$(7, 1)$} & \multirow{2}{*}{$(6, 0)$} & \multirow{2}{*}{$(0, 3)$} & \multirow{2}{*}{$(1, 4)$} & \multirow{2}{*}{$(2, 5)$} & \multirow{2}{*}{$(3, 6)$} & \multirow{2}{*}{$(4, 7)$} & \multirow{2}{*}{$(7, 2)$} & \multirow{2}{*}{$(6, 1)$} & \multirow{2}{*}{$(5, 0)$}
      \tabularnewline Block & & & & & & & & & & & & & & & & \tabularnewline
      \hline
          \multirow{8}{*}{1} & (0, 2) & (0, 3) & (0, 4) & (0, 5) & (0, 6) & (0, 7) & (0, 1) & (0, 0) & (0, 3) & (0, 4) & (0, 5) & (0, 6) & (0, 7) & (0, 2) & (0, 1) & (0, 0)\tabularnewline
          \cline{2-17} \cline{3-17} \cline{4-17} \cline{5-17} \cline{6-17} \cline{7-17} \cline{8-17} \cline{9-17}
          \cline{9-17} \cline{10-17} \cline{5-17} \cline{11-17} \cline{12-17} \cline{13-17} \cline{14-17} \cline{15-17}
          \cline{16-17} \cline{17-17}
          & (1, 2) & (1, 3) & (1, 4) & (1, 5) & (1, 6) & (1, 7) & (1, 1) & (1, 0) & (1, 3) & (1, 4) & (1, 5) & (1, 6) & (1, 7) & (1, 2) & (1, 1) & (1, 0)\tabularnewline
          \cline{2-17} \cline{3-17} \cline{4-17} \cline{5-17} \cline{6-17} \cline{7-17} \cline{8-17} \cline{9-17}
          \cline{9-17} \cline{10-17} \cline{5-17} \cline{11-17} \cline{12-17} \cline{13-17} \cline{14-17} \cline{15-17}
          \cline{16-17} \cline{17-17}
          & (2, 2) & (2, 3) & (2, 4) & (2, 5) & (2, 6) & (2, 7) & (2, 1) & (2, 0) & (2, 3) & (2, 4) & (2, 5) & (2, 6) & (2, 7) & (2, 2) & (2, 1) & (2, 0)\tabularnewline
\hline 
      \end{tabular}
      \par\end{centering}
      \end{adjustbox}
      \medskip{}

      \caption{Correlated modes for each mode of $\mathbf{\Lambda_{5}}$ w.r.t. coefficients belonging to the previous lattices.}\label{tab_L5}
    \end{table*}

    \begin{table*}[h]
      \begin{adjustbox}{width=1.8\columnwidth, center}
      \begin{centering}
      \begin{tabular}{|c|c|c|c|c|c|c|c|c|c|c|c|c|c|c|c|c|}
      \hline Mode / &\multirow{2}{*}{$(0, 4)$} & \multirow{2}{*}{$(1, 5)$} & \multirow{2}{*}{$(2, 6)$} & \multirow{2}{*}{$(3, 7)$} & \multirow{2}{*}{$(7, 3)$} & \multirow{2}{*}{$(6, 2)$} & \multirow{2}{*}{$(5, 1)$} & \multirow{2}{*}{$(4, 0)$} & \multirow{2}{*}{$(0, 5)$} & \multirow{2}{*}{$(1, 6)$} & \multirow{2}{*}{$(2, 7)$} & \multirow{2}{*}{$(7, 4)$} & \multirow{2}{*}{$(6, 3)$} & \multirow{2}{*}{$(5, 2)$} & \multirow{2}{*}{$(4, 1)$} & \multirow{2}{*}{$(3, 0)$}
      \tabularnewline Block & & & & & & & & & & & & & & & & \tabularnewline
      \hline
          \multirow{8}{*}{1} & (0, 4) & (0, 5) & (0, 6) & (0, 7) & (0, 3) & (0, 2) & (0, 1) & (0, 0) & (0, 5) & (0, 6) & (0, 7) & (0, 4) & (0, 3) & (0, 2) & (0, 1) & (0, 0)\tabularnewline
          \cline{2-17} \cline{3-17} \cline{4-17} \cline{5-17} \cline{6-17} \cline{7-17} \cline{8-17} \cline{9-17}
          \cline{9-17} \cline{10-17} \cline{5-17} \cline{11-17} \cline{12-17} \cline{13-17} \cline{14-17} \cline{15-17}
          \cline{16-17} \cline{17-17}
          & (1, 4) & (1, 5) & (1, 6) & (1, 7) & (1, 3) & (1, 2) & (1, 1) & (1, 0) & (1, 5) & (1, 6) & (1, 7) & (1, 4) & (1, 3) & (1, 2) & (1, 1) & (1, 0)\tabularnewline
          \cline{2-17} \cline{3-17} \cline{4-17} \cline{5-17} \cline{6-17} \cline{7-17} \cline{8-17} \cline{9-17}
          \cline{9-17} \cline{10-17} \cline{5-17} \cline{11-17} \cline{12-17} \cline{13-17} \cline{14-17} \cline{15-17}
          \cline{16-17} \cline{17-17}
          & (2, 4) & (2, 5) & (2, 6) & (2, 7) & (2, 3) & (2, 2) & (2, 1) & (2, 0) & (2, 5) & (2, 6) & (2, 7) & (2, 4) & (2, 3) & (2, 2) & (2, 1) & (2, 0)\tabularnewline
\hline 
      \end{tabular}
      \par\end{centering}
      \end{adjustbox}
      \medskip{}

      \caption{Correlated modes for each mode of $\mathbf{\Lambda_{6}}$ w.r.t. coefficients belonging to the previous lattices.}\label{tab_L6}
    \end{table*}

    \begin{table*}[h]
      \begin{adjustbox}{width=1.8\columnwidth, center}
      \begin{centering}
      \begin{tabular}{|c|c|c|c|c|c|c|c|c|c|c|c|c|c|c|c|c|}
      \hline Mode / &\multirow{2}{*}{$(0, 6)$} & \multirow{2}{*}{$(1, 7)$} & \multirow{2}{*}{$(7, 5)$} & \multirow{2}{*}{$(6, 4)$} & \multirow{2}{*}{$(5, 3)$} & \multirow{2}{*}{$(4, 2)$} & \multirow{2}{*}{$(3, 1)$} & \multirow{2}{*}{$(2, 0)$} & \multirow{2}{*}{$(0, 7)$} & \multirow{2}{*}{$(7, 6)$} & \multirow{2}{*}{$(6, 5)$} & \multirow{2}{*}{$(5, 4)$} & \multirow{2}{*}{$(4, 3)$} & \multirow{2}{*}{$(3, 2)$} & \multirow{2}{*}{$(2, 1)$} & \multirow{2}{*}{$(1, 0)$}
      \tabularnewline Block & & & & & & & & & & & & & & & & \tabularnewline
      \hline
          \multirow{8}{*}{1} & (0, 6) & (0, 7) & (0, 5) & (0, 4) & (0, 3) & (0, 2) & (0, 1) & (0, 0) & (0, 7) & (0, 6) & (0, 5) & (0, 4) & (0, 3) & (0, 2) & (0, 1) & (0, 0)\tabularnewline
          \cline{2-17} \cline{3-17} \cline{4-17} \cline{5-17} \cline{6-17} \cline{7-17} \cline{8-17} \cline{9-17}
          \cline{9-17} \cline{10-17} \cline{5-17} \cline{11-17} \cline{12-17} \cline{13-17} \cline{14-17} \cline{15-17}
          \cline{16-17} \cline{17-17}
          & (1, 6) & (1, 7) & (1, 5) & (1, 4) & (1, 3) & (1, 2) & (1, 1) & (1, 0) & (1, 7) & (1, 6) & (1, 5) & (1, 4) & (1, 3) & (1, 2) & (1, 1) & (1, 0)\tabularnewline
          \cline{2-17} \cline{3-17} \cline{4-17} \cline{5-17} \cline{6-17} \cline{7-17} \cline{8-17} \cline{9-17}
          \cline{9-17} \cline{10-17} \cline{5-17} \cline{11-17} \cline{12-17} \cline{13-17} \cline{14-17} \cline{15-17}
          \cline{16-17} \cline{17-17}
          & (2, 6) & (2, 7) & (2, 5) & (2, 4) & (2, 3) & (2, 2) & (2, 1) & (2, 0) & (2, 7) & (2, 6) & (2, 5) & (2, 4) & (2, 3) & (2, 2) & (2, 1) & (2, 0)\tabularnewline
\hline 
      \end{tabular}
      \par\end{centering}
      \end{adjustbox}
      \medskip{}

      \caption{Correlated modes for each mode of $\mathbf{\Lambda_{7}}$ w.r.t. coefficients belonging to the previous lattices.}\label{tab_L7}
    \end{table*}
  %%% -*-BibTeX-*-
  %%% Do NOT edit. File created by BibTeX with style
  %%% ACM-Reference-Format-Journals [18-Jan-2012].

  \bibliographystyle{plain}
  \bibliography{totaleBibDesk}

\begin{thebibliography}{10}

\bibitem{bas:hal-00648057}
P.~Bas, T.~Filler, and T.~Pevny.
\newblock {"Break Our Steganographic System": The Ins and Outs of Organizing
  BOSS}.
\newblock In {\em {INFORMATION HIDING}}, volume 6958/2011 of {\em Lecture Notes
  in Computer Science}, pages 59--70, Czech Republic, September 2011.

\bibitem{bas:WIFS-16}
Patrick Bas.
\newblock {Steganography via Cover-Source Switching}.
\newblock 2016.
\newblock {IEEE Workshop on Information Forensics and Security (WIFS)}.

\bibitem{bas:ICASSP-17}
Patrick Bas.
\newblock {An embedding mechanism for Natural Steganography after
  down-sampling}.
\newblock 2017.
\newblock {IEEE ICASSP}.

\bibitem{cogranne2015ensemble}
R{\'e}mi Cogranne, Vahid Sedighi, Jessica Fridrich, and Tom{\'a}{\v{s}}
  Pevn{\`y}.
\newblock Is ensemble classifier needed for steganalysis in high-dimensional
  feature spaces?
\newblock In {\em Information Forensics and Security (WIFS), 2015 IEEE
  International Workshop on}, pages 1--6. IEEE, 2015.

\bibitem{denemark2015improving}
Tom{\'a}{\v{s}} Denemark and Jessica Fridrich.
\newblock Improving steganographic security by synchronizing the selection
  channel.
\newblock In {\em Proceedings of the 3rd ACM Workshop on Information Hiding and
  Multimedia Security}, pages 5--14. ACM, 2015.

\bibitem{filler2010gibbs}
Tom{\'a}{\v{s}} Filler and Jessica Fridrich.
\newblock Gibbs construction in steganography.
\newblock {\em IEEE Transactions on Information Forensics and Security},
  5(4):705--720, 2010.

\bibitem{filler2011minimizing}
Tomas Filler, Jan Judas, and Jessica Fridrich.
\newblock Minimizing additive distortion in steganography using
  syndrome-trellis codes.
\newblock {\em Information Forensics and Security, IEEE Transactions on},
  6(3):920--935, 2011.

\bibitem{guo2015using}
Linjie Guo, Jiangqun Ni, Wenkang Su, Chengpei Tang, and Yun-Qing Shi.
\newblock Using statistical image model for jpeg steganography: Uniform
  embedding revisited.
\newblock {\em IEEE Transactions on Information Forensics and Security},
  10(12):2669--2680, 2015.

\bibitem{holub2015low}
Vojt{\v{e}}ch Holub and Jessica Fridrich.
\newblock Low-complexity features for jpeg steganalysis using undecimated dct.
\newblock {\em IEEE Transactions on Information Forensics and Security},
  10(2):219--228, 2015.

\bibitem{holub2014universal}
Vojt{\v{e}}ch Holub, Jessica Fridrich, and Tom{\'a}{\v{s}} Denemark.
\newblock Universal distortion function for steganography in an arbitrary
  domain.
\newblock {\em EURASIP Journal on Information Security}, 2014(1):1--13, 2014.

\bibitem{li2015strategy}
Bin Li, Ming Wang, Xiaolong Li, Shunquan Tan, and Jiwu Huang.
\newblock A strategy of clustering modification directions in spatial image
  steganography.
\newblock {\em Information Forensics and Security, IEEE Transactions on},
  10(9):1905--1917, 2015.

\bibitem{li2018defining}
Weixiang Li, Weiming Zhang, Kejiang Chen, Wenbo Zhou, and Nenghai Yu.
\newblock Defining joint distortion for jpeg steganography.
\newblock In {\em Proceedings of the 6th ACM Workshop on Information Hiding and
  Multimedia Security}, pages 5--16. ACM, 2018.

\bibitem{tho2020natural}
Taburet Th{\'e}o, Bas Patrick, Sawaya Wadih, and Jessica Fridrich.
\newblock Natural steganography in jpeg domain with a linear development
  pipeline, 2020.

\bibitem{zhang2016decomposing}
Weiming Zhang, Zhuo Zhang, Lili Zhang, Hanyi Li, and Nenghai Yu.
\newblock Decomposing joint distortion for adaptive steganography.
\newblock {\em IEEE Transactions on Circuits and Systems for Video Technology},
  27(10):2274--2280, 2016.

\end{thebibliography}

\end{document}